\documentclass[11pt,a4paper]{article}

\usepackage{a4wide}
\usepackage{subcaption}
\usepackage[english]{babel}
\usepackage{amsmath}   
\usepackage{graphicx}  % needed for figures
\usepackage{dcolumn}   % needed for some tables
\usepackage{bm}        % for math
\usepackage{amssymb}   % for math
\usepackage{epsfig}
\usepackage{epstopdf}
\usepackage{mdframed}
\usepackage{amsmath}
\usepackage{tikz}
\usetikzlibrary{arrows.meta,calc}
\usepackage{wasysym}
\usepackage{hyperref}
\hypersetup{
    colorlinks=true,
    linkcolor=blue,
    filecolor=magenta,      
    urlcolor=blue,
    citecolor=red
}
\usepackage[toc]{appendix}
\usepackage{color,soul} 
\usepackage{datetime}
\usepackage{physics}
\usepackage{enumerate}
\usepackage{tablefootnote}
\usepackage{multirow}
\usepackage{empheq}
\usepackage{hhline}
\usepackage{esint}
\usepackage[normalem]{ulem}
\usepackage{footmisc}
\usepackage[backend=biber, style=phys, maxbibnames=2, minbibnames=1, doi=true, url=false, eprint=true, isbn=false, biblabel=brackets]{biblatex}
\usepackage{tensor}
\usepackage{hyperref}
\usepackage{csquotes}
\usepackage{orcidlink}

\newcommand{\be}{\begin{equation}}
\newcommand{\ee}{\end{equation}}
\newcommand{\bea}{\begin{eqnarray}}
\newcommand{\eea}{\end{eqnarray}}
\newcommand{\bean}{\begin{eqnarray*}}
\newcommand{\eean}{\end{eqnarray*}}

\newcommand{\FOORT}{\texttt{FOORT} }
\newcommand{\code}[1]{\texttt{#1}}

\newcommand{\nin}{n_\mathrm{init}}
\newcommand{\nmax}{n_\mathrm{max}}
\newcommand{\smax}{s_\mathrm{max}}
\newcommand{\rmd}{\mathrm{d}}

\newcommand{\tc}{t_{\rm c}}
\newcommand{\rc}{r_{\rm c}}
\newcommand{\thc}{\theta_{\rm c}}
\newcommand{\phc}{\phi_{\rm c}}

\begin{document}

% Titlepage
\begin{titlepage}

\numberwithin{equation}{section}
\begin{flushright}
\small
\normalsize
\end{flushright}

\begin{center}

{\LARGE \textbf{FOORT: Flexible Object-Oriented Ray Tracing}}
\\

\medskip

\vspace{1 cm} {\large Daniel R. Mayerson$^{1,2}$ \orcidlink{0000-0003-3595-3330}, Seppe J. Staelens$^{3,2}$ \orcidlink{0000-0002-1262-1600}, Fabio Bacchini$^{4,5,2}$} \orcidlink{0000-0002-7526-8154}\\

\vspace{1cm}
 
 {$^1$ Institute for Theoretical Physics, KU Leuven, Celestijnenlaan 200D, B-3001 Leuven, Belgium}

  {$^2$Leuven Gravity Institute, KU Leuven,
 Celestijnenlaan 200D box 2415, 3001 Leuven, Belgium}

 {$^3$ DAMTP, Centre for Mathematical Sciences, University of Cambridge, Wilberforce Road, Cambridge CB3 0WA, United Kingdom}

 {$^4$ Centre for mathematical Plasma Astrophysics, KU Leuven, Celestijnenlaan 200B, B-3001 Leuven, Belgium}
 
 {$^5$ Royal Belgian Institute for Space Aeronomy, Solar-Terrestrial Centre of Excellence, Ringlaan 3, 1180 Uccle, Belgium}

\vspace{1cm}

 daniel.mayerson@kuleuven.be, ss3033@cam.ac.uk,  fabio.bacchini@kuleuven.be

\vspace{1cm}

\textbf{Abstract}
\end{center}
\noindent 
We present {\tt FOORT} (Flexible Object-Oriented Ray Tracing), an open C++ code for ray-tracing images of black holes and other compact objects in arbitrary spacetimes.
{\tt FOORT} is designed to offer flexibility and can handle horizons, naked singularities and numerical metrics, allowing it to ray-trace numerous different metrics including Kerr and other black holes, as well as horizonless objects such as fuzzballs and boson stars. New metrics, analytical or numerical, can be added with minimal effort, and geodesics are initialized at the camera without reference to conserved quantities, so no special structure of the metric is required.
Besides the intensity observed from a geometrically and optically thin equatorial disk, {\tt FOORT} is set up to track a number of diagnostics along each geodesic, such as the number of equatorial crossings and the radius of closest approach. Its adaptive mesh can refine the image on any of these diagnostics, efficiently resolving narrow features such as higher-order photon rings.
Accompanying Python tools convert the output into images and interferometric visibility amplitudes.
We describe the main features of the code, validate it against analytic results in the Schwarzschild and Kerr spacetimes, demonstrate it on several non-Kerr metrics, and characterize its  scaling with the number of OpenMP threads. {\tt FOORT} is lightweight, requiring only C++17 and OpenMP, and has already been used in a number of studies of black hole and compact-object imaging. The code is publicly available in a GitHub repository.

\vspace{2em}

%%%%%%%%%%%%%%%%%%%%%%%%%%%%%%%%%%%%%%%%%%%%%%%%%%%%%%%%%%%%%%%%%%%%%%
%%%%%%%%%%%%%%%%%%%%%%%%%%%%%%%%%%%%%%%%%%%%%%%%%%%%%%%%%%%%%%%%%%%%%%
% Table of contents
\setcounter{tocdepth}{2}
\tableofcontents

\end{titlepage}
\newpage

\section{Introduction}

When the Event Horizon Telescope (EHT) Collaboration published their studies of the supermassive black holes (SMBHs) M87* \cite{eht_m87} and Sagittarius A* \cite{eht_sgr}, the latter residing at the center of our own Milky Way, they opened up the avenue for a new way to probe black hole physics. Using very long baseline interferometry (VLBI), the EHT Collaboration managed to reconstruct images of these cosmic giants in the millimeter regime, for the first time resolving length scales comparable to the diameter of the event horizon, the one-way membrane from which nothing can escape. While a BH is dark by definition, the plasma in the surrounding accretion disk shines bright, producing the light that is eventually detected. As a consequence, the presence of an SMBH is inferred from the reconstructed image, as a dark central brightness depression associated with the event horizon.

Although the details of the image are heavily based on the astrophysical details of the plasma surrounding the central object, the extent of the dark region can provide an estimate for the mass of the SMBH. In the case of Sagittarius A*, this is consistent with dynamical estimates based on the motion of stars at the center of the Milky Way 
\cite{Do:2019txf, GRAVITY:2021xju}.
However, in order to provide more stringent tests of general relativity (GR) and the Kerr hypothesis, it is essential to identify and analyze imaging observables that are (largely) independent of the uncertainties associated with the details of the plasma \cite{Gralla2021test_GR} (see e.g.\ \cite{Lupsasca:2024wkp} for a pedagogical review). The most promising candidates in this regard are various properties of the photon rings, bright ring-like features in the images that are determined by the underlying spacetime geometry \cite{johnson2020universal}. The photon ring is a consequence of the strong bending of null geodesics near the horizon, and its properties are typically intimately connected to the details of the geometry without depending too much on the details of the emitting plasma --- for example, it has been shown that the GR prediction for the photon ring \emph{shape} can indeed feasibly be measured precisely with near-future VLBI observations \cite{gralla_shape_2020,Paugnat:2022qzy,Lupsasca:2024xhq}.

Large collections of general-relativistic magnetohydrodynamic (GRMHD) plasma simulations are essential to support the data analysis of the EHT studies. This data is typically fed to ray-tracing codes that solve the radiative transfer equations along null geodesics to predict a series of images that we as distant observers would detect. GRMHD simulations are expensive and time consuming, however, such that cheaper alternatives are desirable for theoretical studies in which the precise details of the plasma dynamics are unimportant. Moreover, snapshots of GRMHD simulations contain complicated and variable structures in the plasma, time averaged images of simulations can wash out some of these fluctuations. Simple analytical models, such as that used in \cite{gralla_shape_2020} (and presented below in Sec.~\ref{sec: emission}), can simulate features of time-averaged emission from a geometrically thin accretion disk; this allows for a computationally cheaper method to study what (time-averaged) images of certain BH accretion disks could look like.

\medskip

Most ray-tracing efforts focus on the Kerr geometry, as one expects SMBHs to assume this metric in GR. However, an exciting question to ask is how VLBI images would differ in theories beyond GR, especially when the SMBH is actually not a black hole but rather a different, horizonless exotic compact object (ECO) as could be expected within theories resolving classical black hole paradoxes. Many studies have addressed some aspects of this question, e.g.\ \cite{Staelens:2023photonrings, Carballo-Rubio:2025zwz, bacchini_fuzzball_2021, vincent_imaging_2016, Cunha2015chaotic, Cunha2016chaotic, Rosa2023imaging, gao2025higherorderinnerphotonrings, ayzenberg2023fundamental, Eichhorn:2021iwq, Eichhorn:2022oma}. Exploring the vast zoo of non-Kerr black holes and ECOs is infeasible for full GRMHD simulations --- although some studies have been carried out for e.g.\ boson stars \cite{Olivares:2018abq, Jaramillo:2026ygy}. This underlines the necessity for faster tools, such as ray-tracing codes employing simplified, analytical prescriptions for the emission.

As explained in \cite{Lupsasca:2024wkp}, it is hard to find truly \emph{discriminatory} observables in VLBI black hole imaging, where image predictions from GR can be truly distinguished from those coming from an alternative theory. Properties of the higher-order photon (sub)rings are interesting candidates for observables which may reveal beyond-GR signatures. Specifically, in VLBI imaging, measurements consist of sampling the \textit{visibility}, which is the Fourier transform of the image (see also Sec.~\ref{sec: emission}); the photon rings produce strong universal signatures in the long-baseline regime of the visibility \cite{johnson2020universal}. It is therefore also important to investigate what signatures beyond-GR effects or ECOs would source in the observed visibility in the photon-ring regime. Note that a detection of the $n=1$ photon ring in M87* has been claimed \cite{Broderick:2022tfu}, although this remains debated \cite{Lockhart:2021xel,Lockhart:2022rui}; inclusions of longer baselines by the addition of an antenna in space \cite{Johnson2024BHEX} or on the Moon \cite{Zhao:2026lunar} should enable future upgrades of the EHT to confidently resolve at least the $n=1$ subring, meaning such precise photon-ring observations are plausibly accessible in the near future.

\medskip

A multitude of ray-tracing codes exist, and we refer the reader to e.g.\ \cite{EventHorizonTelescope:2020tst} for a comparison of several ray-tracing codes used in the EHT Collaboration. We highlight several codes here of particular relevance to {\tt FOORT}.

{\tt ipole} \cite{Moscibrodzka:2017lcu, Gelles:2021oyd} and {\tt Blacklight} \cite{White:2022paq} brought adaptive ray tracing to black hole imaging, where more geodesics are integrated in regions of the image that require high resolution. Ref.~\cite{Cardenas-Avendano:2022csp} introduces \texttt{AART}, which focuses on the Kerr geometry and leverages the analytical solutions (see also Ref.~\cite{Dexter:2009fg,Dexter:2016cdk}) to the geodesic equations to improve accuracy, while also boasting a layered mesh that allows for high resolution in the series of lensing bands, where the photon rings reside. This enables the user to better resolve the photon rings, which become exponentially narrower as $n$ increases. Additionally, \texttt{AART} goes beyond the GLM prescription (see Sec.~\ref{sec: emission}) for the optically thin accretion disk, and can sample noise realizations on top of this to mimic a more dynamical disk, coming closer to standard GRMHD simulations.

Various ray-tracers also allow for non-Kerr spacetimes. {\tt GYOTO} \cite{Vincent:2011wz} is a well-established code for analytical and numerical spacetimes, with a range of emission sources including thin disks and orbiting stars, and has been used e.g.\ for boson stars \cite{vincent_imaging_2016}.
\texttt{PyHole} \cite{Cunha2016chaotic} was created to ray-trace BHs with scalar hair and boson stars.
\texttt{RAPTOR} \cite{bronzwaer2018raptor} handles analytical and numerical spacetimes, runs on both CPUs and GPUs, and implements time-dependent (``slow-light'') radiative transfer, in which the plasma evolves while the light propagates. \texttt{BHOSS} \cite{Younsi:2019iee} has been used, among others, to compare accreting boson stars with classical BHs \cite{Olivares:2018abq}.
 A number of other ray-tracers or geodesic integrators allow for arbitrary spacetimes, e.g.\ \texttt{Arcmancer} \cite{Pihajoki:2018ihj}, \texttt{Mahakala} \cite{Sharma:2023nbk}, \texttt{FANTASY} \cite{Christian:2020xrp}, and \texttt{Gradus.jl} \cite{Baker:2025ifr}; recently, the Python code {\tt TARTARUS} \cite{Bambague:2026jey}, which uses just-in-time compilation for speed. \texttt{Mahakala}, \texttt{FANTASY}, and \texttt{Gradus.jl} compute the Christoffel symbols by automatic differentiation, whereas {\tt FOORT} uses finite differences (see Sec.~\ref{sec: basic}).

\medskip

In this work, we introduce a new ray-tracing code, {\tt FOORT} \cite{Mayerson_FOORT_Flexible_Object-Oriented_2026}, with different objectives in mind, which distinguish it from the plethora of other ray-tracers\footnote{The code is publicly available at \url{https://github.com/drmayerson/FOORT}.}:
\begin{itemize}
    \item Primarily, \FOORT is focused on modularity and adaptability. The code is written in C++ and allows the user to extend the code with minimal effort to add new spacetime geometries and additional diagnostics. 
    \item {\tt FOORT} is specifically designed for (but not limited to) studying non-Kerr geometries, in GR and beyond, both with and without horizons. A range of metrics are already implemented, ranging from classical BHs to (horizonless) fuzzballs and boson stars --- see App.~\ref{app: metrics}. While several other codes can handle non-Kerr geometries, {\tt FOORT}'s built-in framework can easily handle horizons, naked singularities of arbitrary codimension, horizonless objects, and numerical metrics with its built-in termination conditions. In addition, the light rays originating at the camera are initialized without any reference to conserved quantities, making the camera metric-agnostic.
    \item \FOORT also features a flexible adaptive mesh refinement (AMR) scheme that allows for the resolution of small image features. Importantly, the adaptive mesh can be refined on the basis of \emph{any} diagnostic variable that is tracked during integration, not just the observed intensity. Many such diagnostics are already built in {\tt FOORT}, including a four-color screen, equatorial passes, and closest-approach radius.
    \item \FOORT is designed to be user-friendly, keeping a possible user base in mind that may not be overly familiar with (ray-tracing) codes. \FOORT is also light-weight and self-contained, requiring only C++17 and OpenMP and no external libraries. It includes Python tools to facilitate converting raw \FOORT output to images as well as visibility amplitudes.
\end{itemize}

\FOORT has already been used in several works. Its first use case was to study whether BHs that deviate from Kerr can lead to observable differences in their photon rings; more precisely in their shape and the Lyapunov exponent \cite{Staelens:2023photonrings}. It has also been used to image fuzzballs \cite{mayerson2023observational}, which have been shown to cast emergent shadows due to their complex microstructure, despite the absence of an event horizon \cite{bacchini_fuzzball_2021}. \FOORT has been used to study the shadows of scalarized BHs, including non-trivial Ricci couplings \cite{Fernandes2024shadows}, and BHs in astrophysical environments \cite{Fernandes:2025osu}. More recently, it has been used to investigate the influence of a non-minimal coupling between the curvature and the electromagnetic field \cite{Carballo-Rubio:2025zwz} and to image ultracompact boson stars \cite{Marks:2025jpt, Staelens:2026rzb} and wormholes \cite{Mehta:2025vis}.

\medskip

The rest of this paper is structured as follows. In Sec.~\ref{sec: basic}, we introduce the essential geodesic equations and the strategy that {\tt FOORT} takes to integrate them, and we show tests of {\tt FOORT}'s performance, precision and scaling. Sec.~\ref{sec: specific} discusses the more {\tt FOORT}-specific features of its metric-agnostic geodesic initialization procedures, the different features included to analyze the ultimate output, the adaptive mesh algorithm, and its emission model that can be used to simulate GRMHD images and interferometric visibility amplitudes. Then, in Sec.~\ref{sec: demos}, we provide several explicit demonstrations of images produced by {\tt FOORT} for various metrics. Finally, we provide our conclusions in Sec.~\ref{sec: conclusion}. App.~\ref{app: metrics} provides details of all the different metrics currently implemented in {\tt FOORT}, and App.~\ref{app: fluid velocity} discusses further details of the simulated accretion disk properties used in the emission model.

Note that {\tt FOORT} includes extensive documentation on all implemented metrics, diagnostics, terminations, outputs, as well as its Python library {\tt pyFOORT} that serves to facilitate creating images out of {\tt FOORT} raw outputs. For technical or practical details on how to use any particular feature discussed in this paper, we refer to this documentation.
Unless otherwise specified, we set $G = c =1$.

%%%%%%%%%%%%%%%%%%%%%%%%%%%%%%%%%%%%%%%%%%%%%%
%%%%%%%%%%%%%%%%%%%%%%%%%%%%%%%%%%%%%%%%%%%%%%
%%%%%%%%%%%%%%%%%%%%%%%%%%%%%%%%%%%%%%%%%%%%%%
\section{Basic Ray Tracing Code Features}\label{sec: basic}

The goal of a ray-tracing code in GR is to determine what a spacetime exhibiting a region of strong gravity --- typically a BH or another compact object --- looks like for an observer located at a large distance. Typically, the spacetime is assumed to be asymptotically flat, such that the observer is located in a region where the spacetime is approximately (spherical) Minkowski. To determine what the distant observer sees, a collection of null geodesics is traced back from its position in the direction of the central region.

The geodesic equations are cast in a first-order system:
\begin{align}
    \frac{\rmd x^\mu}{\rmd\tau} & = u^\mu, \label{eq: geodesic equation x}\\
    \frac{\rmd u^\mu}{\rmd\tau} & = -\Gamma^\mu_{\lambda\sigma} u^\lambda u^\sigma, \label{eq: geodesic equation u}
\end{align}
where $\tau$ is an affine parameter and $\Gamma^\mu_{\lambda\sigma}$ are the usual Christoffel symbols. Greek indices run from 0 to 3 and correspond to the coordinates in which the metric is given. Note that certain features of {\tt FOORT} implicitly assume a spherical-type coordinate system with $(x^2,x^3)=(\theta,\phi)$ with $\theta$ ranging between $0$ and $\pi$ while $\phi$ ranges between $0$ and $2\pi$; in particular, this assumption can be seen in the geodesic initialization and the adaptive step size calculation below.

\subsection{Geodesic Integration}

The user can specify a position for the distant observer; the null geodesics are then initialized at this observer's position with an appropriate initial velocity to simulate a pinhole camera of an appropriate angular aperture at the observer's location. This velocity is past-pointing as we are integrating the geodesic \emph{backwards} in time starting at the observer (see Sec~\ref{sec: viewscreen} below for further details).

\FOORT implements standard integration schemes for numerically solving the geodesic equations~\eqref{eq: geodesic equation x}--\eqref{eq: geodesic equation u}. At the moment, the code supports a fourth order Runge--Kutta (RK4) scheme and a second-order velocity Verlet algorithm.
By default, the Verlet scheme is selected, which advances the geodesic equations as follows \cite{Dolence_2009}:
\begin{align}
    x^\mu_{n+1} & = x^\mu_n +  \Delta \tau \,u^\mu_n + \frac{(\Delta \tau)^2}{2} \left(\frac{\rmd u^\mu}{\rmd \tau}\right)_n \,, \\
    u^\mu_{n+1, p} & = u^\mu_n + \Delta \tau \left(\frac{\rmd u^\mu}{\rmd \tau}\right)_n\,, \\
    \left(\frac{\rmd u^\mu}{\rmd \tau}\right)_{n+1} & = -\Gamma_{\lambda\sigma}^\mu (x^\mu_{n+1}) \,u^\lambda_{n+1,p}\,u^\sigma_{n+1,p}\,, \label{eq: iter accel}\\
    u^\mu_{n+1} & = u^\mu_n + \frac{\Delta \tau}{2}\left[\left(\frac{\rmd u^\mu}{\rmd \tau}\right)_n + \left(\frac{\rmd u^\mu}{\rmd \tau}\right)_{n+1} \right] \,. \label{eq: iter vel}
\end{align}
In this form, the scheme is explicit. However, steps~\eqref{eq: iter accel}--\eqref{eq: iter vel} can be iterated until the change in the velocity vector $u_{n+1}^\mu$ is below some predefined tolerance, making the scheme semi-implicit: upon iteration, $u_{n+1,p}$ is replaced by the most recent $u_{n+1}$. By default, this iteration loop is turned off, but if activated by the user it can improve the accuracy of the integration.

The advantage of the Verlet scheme is that the Christoffel symbols only need to be evaluated twice per step, at $x^\mu_n$ and $x^\mu_{n+1}$, irrespective of this final iteration. This can offer a significant speedup for more complex spacetime metrics, enabling a faster visualization process. However,  it should be noted that RK4 gives more accurate results, especially for relatively simple spacetimes, like Kerr: we thus recommend the user to consider both integrators for production runs. On the other hand, it has been shown that symplectic integrators, unlike the ones considered here, possess better geometric properties for long geodesic integrations (see, e.g.,~Refs.~\cite{Bacchini_2018, hairer2006geometric}). Such symplectic integrators can be included in future versions of {\tt FOORT}.
Note that the calculation of the Christoffel symbols in the geodesic equations requires spacetime derivatives of the metric, which are calculated numerically using second-order finite differencing:
\begin{equation}\label{eq: finite diff Christoffel}
\partial_\rho g_{\mu\nu}(x^\sigma) = \frac{g_{\mu\nu}(x^\sigma+h\, \delta^\sigma_\rho) - g_{\mu\nu}(x^\sigma-h\, \delta^\sigma_\rho)}{2h},
\end{equation}
where $h$ is a parameter that can be set in the {\tt FOORT} configuration, defaulting to $h=10^{-7}$.

Furthermore, \FOORT also makes use of the adaptive step-size routine introduced in Refs.~\cite{Dolence_2009, noble2007simulating} and used in other ray tracing codes such as Ref.~\cite{bronzwaer2018raptor}. The step-size $\Delta \tau$ is determined by\footnote{The code uses the velocity of the first coordinate. If using the logarithmic radius $\log r$, then effectively $u^r$ is replaced by $u^r/r$ in these equations. Note that this changes the step from $\sim\epsilon$ to $\sim \epsilon r$, which is what makes the $\log r$ approach cheap when integrating at large radii.} 
\begin{equation}\label{eq: adaptive step size}
    \Delta \tau = \epsilon \left[|\Delta \tau_1|^{-1} + |\Delta \tau_2|^{-1} + |\Delta \tau_3|^{-1}\right]^{-1}\,,
\end{equation}
where
\begin{align}
    \Delta \tau_1 & = (|u^r| + \delta)^{-1}, \\
    \Delta \tau_2 & = \min\left(\theta, \pi-\theta\right) (|u^\theta| + \delta)^{-1}, \\
    \Delta \tau_3 & = (|u^\phi| + \delta)^{-1}\,.
\end{align}
In this approach, $\delta$ is a small positive parameter --- set to $10^{-20}$ --- that prevents division by 0, whereas $\epsilon$ determines the overall scale of the step size. {\tt FOORT} also has a configuration parameter setting a minimum step-size, which defaults to $10^{-12}$. The above routine improves accuracy and efficiency in regions where the numerical integration becomes challenging, like in the vicinity of the event horizon and the poles. For numerical efficiency, most implemented metrics support logarithmic rescaling of the radial coordinate, i.e.\ evolving $\Lambda = \log r$ instead of $r$.

\subsection{Tests \& performance}
To test the code's performance on parallel architectures, we carry out a standard geodesic calculation for Minkowski (flat), Schwarzschild, and Kerr (with spin $a=0.99M$) spacetimes. We employ a uniform mesh with $65536$ pixels, and in all cases we set the camera at $r=1000M$, $\theta=0.2967$ ($\sim17^\circ$), $\phi=0$, with a screen size of $15M\times15M$. Numerical integration uses the Verlet scheme, with \texttt{StepSize = 0.03} and \texttt{SmallestPossibleStepsize = 1e-7}. This setting provides a scenario that is relatively imbalanced for parallel calculations: when a BH is present, the shadow covers a large fraction of the viewscreen, implying that many (infalling) geodesics will be integrated for a much shorter time (i.e.\ until they reach the horizon) than those that hover just around the shadow's edge. Geodesics passing sufficiently far from the BH will also take a much shorter time to be integrated. In essence, a handful of light rays that are integrated for a long time are spread across a few threads, and are therefore driving the overall code's performance.

\begin{figure}[htbp]\centering
 \includegraphics[width=0.65\textwidth]{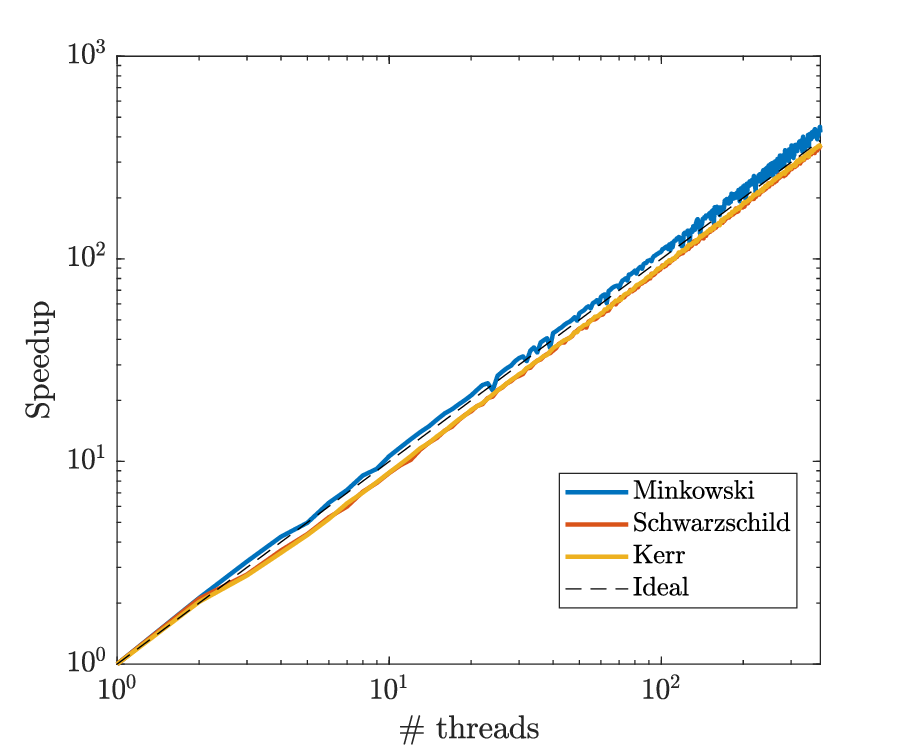}
 \caption{Scaling test for \FOORT up to 384 threads, tracing 65536 pixels on a uniform grid in Minkowski (flat), Schwarzschild, and Kerr ($a=0.99M$) spacetimes. The scaling remains very close to the ideal line when ray tracing BHs, even though a large part of the screen is covered by black pixels (infalling geodesics) that add imbalance to the scaling.}
 \label{fig:scaling}
\end{figure}

We conduct this calculation on an increasing number of OMP threads and record the runtime. The results are shown in Fig.~\ref{fig:scaling}, where we report the code's speedup from the baseline (i.e.\ the runtime on 1 OMP thread) up to 384 threads. We observe that the code performs fairly well throughout the scaling, in spite of the imbalance brought about by this choice of metric. At 384 threads, the measured speedup deviates from the ideal scaling by $\sim2\%$.

\begin{figure}
    \centering
    \includegraphics[width=0.6\linewidth]{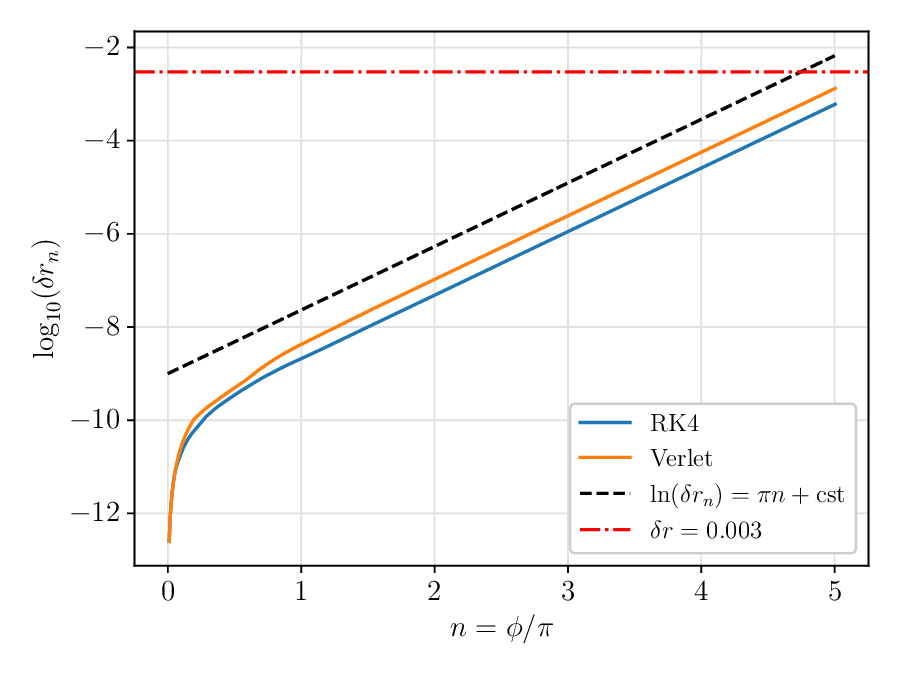}
    \caption{Deviation of the radial coordinate of a bound, circular geodesic initialized at $x^\mu = (0, 3M, \pi/2, 0)$ with velocity vector tangent to $\partial_t$ and $\partial_\phi$. Theoretically, the geodesic should remain at $r=3M$, but numerical errors quickly perturb it, after which the growth in the deviation follows the characteristic exponential divergence as function of the number of half-orbits $n$, set by the Lyapunov exponent $\gamma_L = \pi$. Both integration schemes perform comparatively well, with a fractional error below 0.1\% for the first 5 half-orbits.}
    \label{fig: lyapunov}
\end{figure}

Ultracompact spacetimes contain bound null geodesics that neither escape nor fall into a potential horizon. These bound geodesics lead to photon rings: thin, nearly circular features that are expected to contribute significantly to images of BHs and dominate on long interferometric baselines \cite{johnson2020universal} --- see e.g.\ Ref.~\cite{Lupsasca:2024wkp} for a recent review.
The photon rings are caused by nearly bound geodesics that approximate the bound ones: the radial coordinate of the nearly bound geodesic exhibits a characteristic exponential divergence away from the radius $r_b$ of the bound geodesic, determined by the Lyapunov exponent $\gamma_L$.
We have used {\tt FOORT} to integrate the bound circular geodesic at $r=3M, \theta = \pi/2$ in Schwarzschild spacetime. Due to discretization errors, the geodesic quickly starts deviating from $r=3M$, thus becoming nearly bound. Figure~\ref{fig: lyapunov} shows the value of the radial coordinate as a function of the number of half-orbits (corresponding to $\phi$ changing by $\pi$), showing that the geodesic indeed exponentially diverges away from $r=3M$ at the correct rate set by the Lyapunov exponent for Schwarzschild spacetime $\gamma_L = \pi$ \cite{johnson2020universal}. This test demonstrates that {\tt FOORT} is able to reproduce the characteristic behavior of nearly bound orbits.

\begin{figure}
    \centering
    \includegraphics[width=0.6\linewidth]{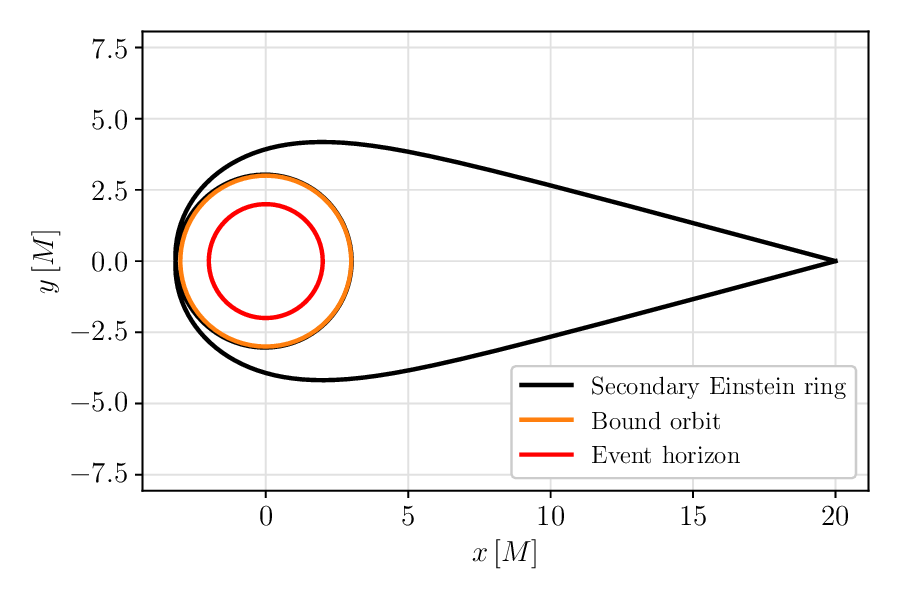}
    \caption{Visualization of the null geodesics in Schwarzschild spacetime relevant for testing the code's performance. Black: a geodesic creating a secondary Einstein ring at $r= 20 M$. Orange: the bound orbit at $r = 3M$. The red circle represents the event horizon at $r = 2M$.}
    \label{fig: geodesics}
\end{figure}

As a second test, we use \FOORT to integrate the geodesic shown in black in Fig.~\ref{fig: geodesics}. This geodesic starts at an observer located at $(r=r_0 = 20M, 
\theta = \pi/2, \phi = 0)$, to which it returns after orbiting the black hole twice, giving rise to a secondary Einstein ring. The coordinate time elapsed for this geodesic is calculated\footnote{These reference values are obtained by high-precision evaluation of the analytic orbit integrals.} to be $\Delta t^* \approx 100.792044145 M$ (see e.g.\ \cite{Bacchini_2018, muller2008einstein}). The initial velocity vector of the geodesic is given by 
\begin{equation}
    \boldsymbol{v} = \left(\frac{1}{\sqrt{-g_{tt}(r_0)}}, -\frac{\cos \Xi}{\sqrt{g_{rr}(r_0)}}, 0, \frac{\sin\Xi}{r_0}  \right)\,,
\end{equation}
where $\Xi = 0.2490496437949734$. We stop the integration of the geodesic once it passes $r=r_0$ again, using the boundary sphere termination (see Sec.~\ref{sec: output and analysis}), and interpolate onto that sphere (in postprocessing) to obtain the final coordinates $\boldsymbol{x}_f$ of the geodesic. We then compare the difference between this position and the expected values using the error measure
\begin{equation}\label{eq: distance metric einstein ring}
    \Delta l = \sqrt{(t_f - t_f^*)^2 + r_f^2 + (r_f^*)^2 - 2r_f r_f^*\, \cos(\phi_f - \phi_f^*)}\,,
\end{equation}
where $(t_f^*, r_f^*, \phi_f^*) = (\Delta t^*, r_0, 0)$ are the analytic values mentioned before. 

\begin{figure}
    \centering
    \includegraphics[width=0.6\linewidth]{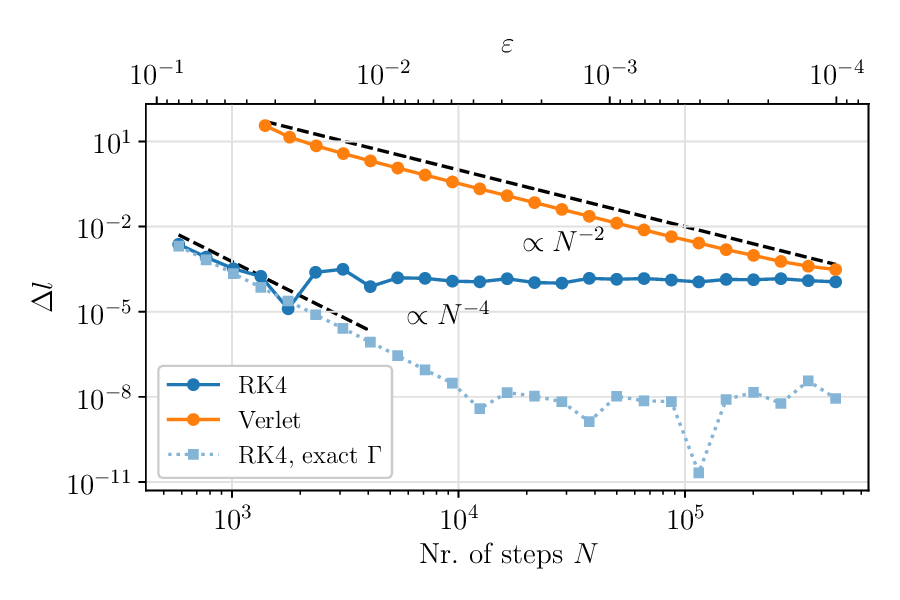}
    \caption{Distance metric (\ref{eq: distance metric einstein ring}) as function of the number of integration steps $N$ for the black geodesic in Fig. \ref{fig: geodesics}. The number of integration steps is controlled by the parameter $\epsilon$ in (\ref{eq: adaptive step size}), which is approximately linearly related to $N^{-1}$, as illustrated by the top axis. The Verlet method (with the velocity iteration turned off here) exhibits second-order convergence, and the RK4 integrator flattens out at $\mathcal{O}(10^{-4})$ due to the finite-difference stencil \eqref{eq: finite diff Christoffel}. Using analytic expressions for the Schwarzschild Christoffel symbols maintains fourth-order convergence for longer.}
    \label{fig: convergence}
\end{figure}

Figure~\ref{fig: convergence} shows the values of this metric as a function of the number of integration steps $N$. The number of steps is not directly controlled, but rather a consequence of the value of the parameter $\epsilon$ in the adaptive step size routine (\ref{eq: adaptive step size}): over the range plotted, $N \propto \epsilon^{-1}$ to good approximation. In this test, RK4 clearly outperforms Verlet, which displays the expected second-order convergence. RK4 is fourth-order accurate, which is seen for the smallest $N$, but convergence stalls and the error measure \eqref{eq: distance metric einstein ring} flattens at $\mathcal{O}(10^{-4})$: this is a consequence of the finite-difference stencil \eqref{eq: finite diff Christoffel}. Indeed, using analytic expressions for the Schwarzschild Christoffel symbols extends the regime of fourth-order convergence until a plateau at $\mathcal{O}(10^{-8})$ is reached, set by round-off error.
As noted before, for more complex spacetimes no analytic Christoffel symbols may be available, and evaluating the metric can be more expensive. As the Verlet integrator is cheaper per step, it usually suffices in initial, exploratory simulations where the results need not be extremely precise.

%%%%%%%%%%%%%%%%%%%%%%%%%%%%%%%%%%%%%%%%%%%%%%
%%%%%%%%%%%%%%%%%%%%%%%%%%%%%%%%%%%%%%%%%%%%%%
%%%%%%%%%%%%%%%%%%%%%%%%%%%%%%%%%%%%%%%%%%%%%%
\section{Specific \FOORT Code Features} \label{sec: specific}

\FOORT is designed with flexibility in mind: in particular, it is designed to be as straightforward as possible to add new metrics, diagnostics, and terminations. Furthermore, the code is made efficient through the use of an adaptive mesh, enabling the image to obtain high resolution in the regions of interest. The current implementation of {\tt FOORT} already contains a number of metrics, which are listed and briefly described in App.~\ref{app: metrics}.

\subsection{Geodesic initialization}\label{sec: viewscreen}

Ray tracers specializing to Kerr often construct the initial photon momenta at the camera using
the conserved quantities of the spacetime. {\tt FOORT}, instead, is designed to initialize its geodesics in a metric-agnostic way. We assume only that the metric is asymptotically flat in coordinates $(t,r,\theta,\phi)$ that are (asymptotically) spherical, and that $\partial_t$ is timelike at the observer location (so that a static camera position is physically sensible). The camera is then placed at a location $(\rc,\thc,\phc)$, chosen by the user, and points at the origin. The construction of the photon initial momenta proceeds in two steps: we first initialize the camera and light rays in flat space, and then map the resulting flat-space momentum to the curved spacetime using a particular local vielbein.

\subsubsection{Flat-space camera and view screen}\label{sec: flat space camera}
 
If $(r_0,\theta_0,\phi_0)$ are ordinary spherical coordinates in flat
space (related to usual flat space Cartesian coordinates $(x_0,y_0,z_0)$ in the canonical way), we can introduce Cartesian camera coordinates $(\tilde x,\tilde
y,\tilde z)$ with their origin at the observer location $(r_0,\theta_0,\phi_0)=(\rc,\thc,\phc)$, with $\tilde z$ pointing away from
the origin (opposite to the viewing direction) and $\tilde y$ aligned with
the projection of the original $z_0$-axis onto the plane orthogonal to the
line of sight. These follow from the original Cartesian coordinates by a
translation to the camera position and two rotations; the resulting axes
are simply
\begin{equation}
  \hat e_{\tilde x} = \hat{\phi}_0\,,\qquad
  \hat e_{\tilde y} = -\hat{\theta}_0\,,\qquad
  \hat e_{\tilde z} = \hat{r}_0\,,
  \label{eq:camframe}
\end{equation}
evaluated at the camera location (and using hats to denote unit vectors in a given direction). For example, a camera on the positive $x_0$-axis
($\thc=\pi/2$, $\phc=0$) has $(\tilde x,\tilde y,\tilde z)=(y_0,z_0,x_0-\rc)$.

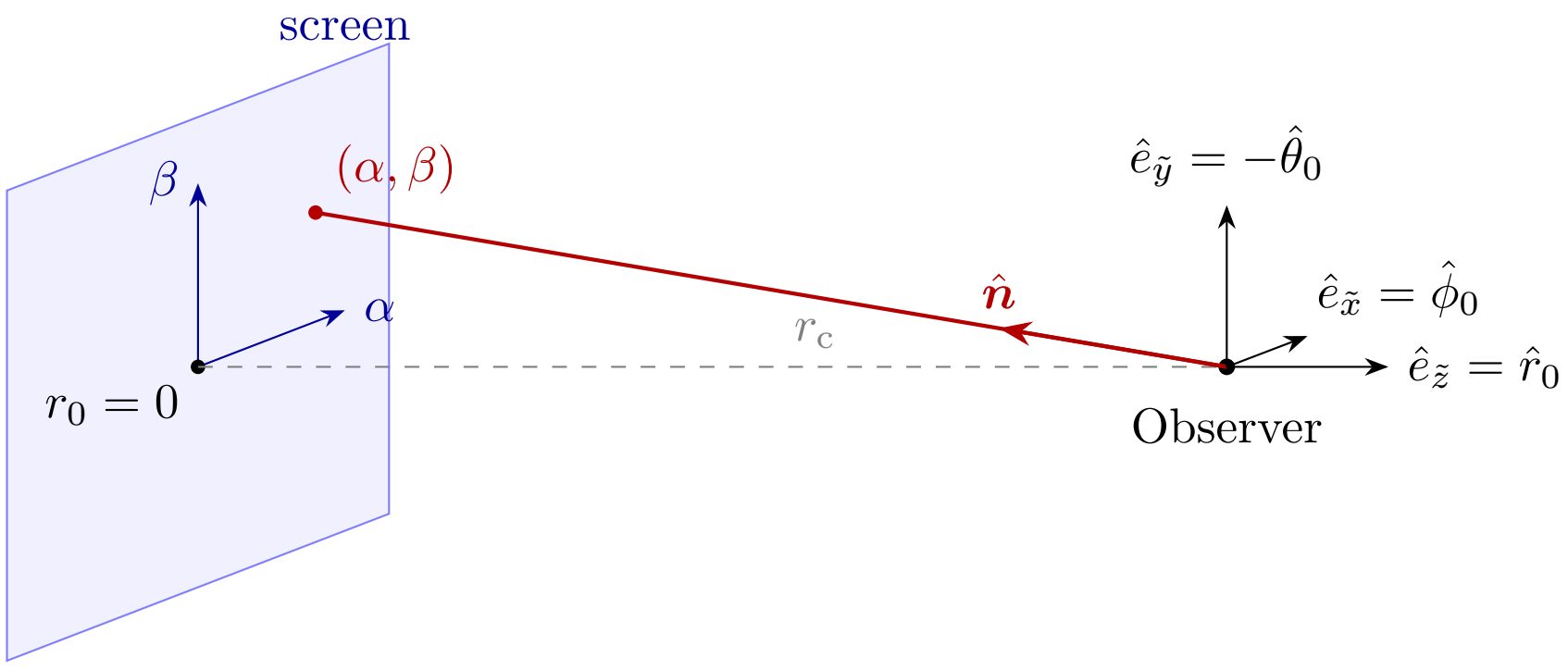
\begin{figure}[t]
\centering
\begin{tikzpicture}[scale=1.05,>=Stealth]
  % screen plane (oblique view)
  \coordinate (O) at (0,0);
  \filldraw[fill=blue!6,draw=blue!50] (-1.3,-2.0) -- (1.3,-1.0) -- (1.3,2.2) -- (-1.3,1.2) -- cycle;
  \node[blue!60!black] at (1.0,2.3) {\small screen};
  % screen axes
  \draw[->,blue!60!black] (O) -- (1.0,0.385) node[right,font=\small] {$\alpha$};
  \draw[->,blue!60!black] (O) -- (0,1.25) node[left,font=\small] {$\beta$};
  \fill (O) circle (1.4pt) node[below left,font=\small] {$r_0=0$};
  % camera
  \coordinate (C) at (7,0);
  \fill (C) circle (1.6pt);
  \node[below,font=\small] at ($(C)+(0,-0.15)$) {Observer};
  % line of sight
  \draw[dashed,gray] (O) -- (C);
  \node[gray,font=\small,above] at (4.2,0) {$\rc$};
  % pixel and ray
  \coordinate (P) at (0.8,1.05);
  \fill[red!70!black] (P) circle (1.4pt) node[above right,font=\small] {$(\alpha,\beta)$};
  \draw[red!70!black,thick] (C) -- (P);
  \draw[->,red!70!black,thick] (C) -- ($(C)!0.25!(P)$) node[above,font=\small,pos=1] {$\hat{\boldsymbol{n}}$};
  % camera axes
  \draw[->] (C) -- ($(C)+(1.1,0)$) node[right,font=\small] {$\hat e_{\tilde z}=\hat r_0$};
  \draw[->] (C) -- ($(C)+(0,1.1)$) node[above,font=\small] {$\hat e_{\tilde y}=-\hat\theta_0$};
  \draw[->] (C) -- ($(C)+(0.55,0.21)$) node[above right,font=\small,xshift=-2pt] {$\hat e_{\tilde x}=\hat\phi_0$};
\end{tikzpicture}
\caption{The camera sits at $(r_0,\theta_0,\phi_0)=(\rc,\thc,\phc)$ and points at the origin.
The pixel $(\alpha,\beta)$ on the screen through the origin defines the
initial direction $\hat n$ of the (backwards-traced) ray.}
\label{fig:screen}
\end{figure}

We can then consider the viewscreen as the plane through the origin $r_0=0$, orthogonal to the line of sight, with Cartesian coordinates $(\alpha,\beta)$ along $\hat e_{\tilde
x}$ and $\hat e_{\tilde y}$; see Fig.~\ref{fig:screen}. The pixel $(\alpha,\beta)$ is assigned the (flat-space) ray that runs from the camera to the point $(\tilde x,\tilde y,\tilde z)=(\alpha,\beta,-\rc)$,
so that the camera acts as a pinhole camera with the screen at distance
$\rc$.\footnote{Equivalently, $(\alpha,\beta)$ are stereographic
coordinates: they are the stereographic projection, from the camera, of
the point where the ray crosses the sphere that has the camera and the
origin as antipodal points (or, alternatively, the sphere of radius $\rc$
centered on the origin).} This ray running from the camera towards the $(\alpha,\beta)$ pixel on the viewscreen has unit spatial direction:
\begin{equation}
  \hat{\boldsymbol{n}} = (\hat n_{\tilde x},\hat n_{\tilde y},\hat n_{\tilde z})=  \frac{1}{\sqrt{\rc^2+\alpha^2+\beta^2}}\,(\alpha,\beta,-\rc)\,.
  \label{eq:ndir}
\end{equation}
We trace geodesics backwards in time, so we take the photon momentum to
be past-directed and pointing into the screen. Using
\eqref{eq:camframe} and \eqref{eq:ndir}, the components of the photon's momentum in the flat-space spherical coordinates $(t_0,r_0,\theta_0,\phi_0)$ are
\begin{equation}
  p^{a} = (p^{t_0},p^{r_0},p^{\theta_0},p^{\phi_0}) = \frac{\tilde E}{\sqrt{\rc^2+\alpha^2+\beta^2}}
  \left(-\sqrt{\rc^2+\alpha^2+\beta^2},\;
    -\rc,\;
    -\frac{\beta}{\rc},\;
    \frac{\alpha}{\rc\sin\thc}\right),
  \label{eq:pflat}
\end{equation}
which is null with respect to the flat metric at the camera
\be \label{eq:etasph}
\eta_{ab} = \mathrm{diag}\left(-1,\,1,\,\rc^2,\,\rc^2\sin^2\thc\right),\ee
for any normalization $\tilde E>0$.
 
\subsubsection{Local vielbein and curved-space initial momentum}

We construct a vielbein $e^\mu_{a}$, normalized to the flat
spherical metric~\eqref{eq:etasph}, to carry \eqref{eq:pflat} over to curved spacetime. We use a Gram--Schmidt orthogonalization procedure in the order $(t,\phi,r,\theta)$, with seed vectors: 
\begin{equation}
  \partial_t\,,\qquad \partial_\phi\,,\qquad
  \nabla r = g^{r\mu}\partial_\mu\,,\qquad
  \nabla\theta = g^{\theta\mu}\partial_\mu\,.
\end{equation}
Using the gradients $\nabla r$ and $\nabla\theta$, rather than $\partial_r$
and $\partial_\theta$, is what makes the construction robust: since
$g(\nabla r,X)=X^r$ for any vector $X$, the gradient $\nabla r$ is
automatically orthogonal to $\partial_t$, $\partial_\phi$, and
$\partial_\theta$, and similarly for $\nabla\theta$. The Gram--Schmidt
procedure then collapses to the explicit vielbein
\begin{equation}
\begin{aligned}
  e_{t_0} &= \frac{\partial_t}{\sqrt{-g_{tt}}}\,, &
  e_{\phi_0} &= \rc\sin\thc\,\frac{\partial_\phi - (g_{t\phi}/g_{tt})\,\partial_t}
                    {\sqrt{g_{\phi\phi}-g_{t\phi}^2/g_{tt}}}\,,\\[4pt]
  e_{r_0} &= \frac{\nabla r}{\sqrt{g^{rr}}}\,, &
  e_{\theta_0} &= \rc\,\frac{\nabla\theta - (g^{r\theta}/g^{rr})\,\nabla r}
                    {\sqrt{g^{\theta\theta}-(g^{r\theta})^2/g^{rr}}}\,.
\end{aligned}
\label{eq:tetrad}
\end{equation}
This vielbein requires
no symmetry assumptions or particular structure of the metric.  Physically, $e_{t_0}$ is the four-velocity of
a static observer, so the image is the one seen by a camera at rest with
respect to the asymptotic frame.
We evaluate and use our vielbein at the camera location $x^\mu = (\tc,\rc,\thc,\phc)$.

The initial momentum of a photon in curved-space coordinates is
\begin{equation}
  p^\mu = e^\mu_{a}\, p^{a}\,,
  \label{eq:pcurved}
\end{equation}
with $p^{a}$ given by Eq.~\eqref{eq:pflat}. Since the factors of $\rc$ in $\eta_{ab}$ cancel between $p^a$ and the
vielbein, Eq.~\eqref{eq:pcurved} can equivalently be written in terms of the
unit vectors $\hat e_{\theta_0}\equiv e_{\theta_0}/\rc$ and
$\hat e_{\phi_0}\equiv e_{\phi_0}/(\rc\sin\thc)$ as
\begin{equation}
  p^\mu = \tilde E\left[-e_{t_0} + \frac{\alpha\,\hat e_{\phi_0}
  - \beta\,\hat e_{\theta_0} - \rc\, e_{r_0}}{\sqrt{\rc^2+\alpha^2+\beta^2}}\right]^\mu .
  \label{eq:pinhole}
\end{equation}
The construction thus amounts to a pinhole camera in the rest frame of the
static observer, with the screen at distance $\rc$: the pixel
$(\alpha,\beta)$ corresponds to a ray at angle $\Theta$ from the line of
sight, with $\tan\Theta=\sqrt{\alpha^2+\beta^2}/\rc$, and with $\alpha$ and
$\beta$ measured along $\hat e_{\phi_0}$ and $-\hat e_{\theta_0}$
respectively.\footnote{The auxiliary flat space only serves to label null directions at
the camera by screen coordinates $(\alpha,\beta)$; it does not identify
points of flat and curved space. The camera is placed at the curved-space
coordinates $(\rc,\thc,\phc)$, and every ray is an exact null geodesic
through it. Note that $\rc$ also sets the camera--screen distance in Eq.~\eqref{eq:pinhole}; this fixes the scale of
$(\alpha,\beta)$.}

The parameter $\tilde
E$ is the photon energy measured by the (static) observer. The conserved
energy associated with $\partial_t$ (for the physical, future-directed
photon with momentum $-p^\mu$) is
\begin{equation}
  E = p_t = \sqrt{-g_{tt}}\;\tilde E\,,
\end{equation}
and note that at large $\rc$,
$\tilde E = E + \mathcal{O}(\rc^{-1})$. We simply set $E = 1$ for all geodesics in {\tt FOORT}.

This choice of vielbein gives sensible results both for metrics in Boyer--Lindquist-type coordinates and for metrics with many off-diagonal components, such as Kerr in Kerr--Schild coordinates; in the latter case $\partial_r$ is not orthogonal to
$\partial_t$, and is in fact null everywhere ($g_{rr}=0$),
whereas $g(\nabla r,\nabla r)=g^{rr}$ is positive precisely outside the
horizon. For large $\rc$ in the Kerr spacetime, the
physical photon momentum $k_\mu=-p_\mu$ satisfies $k_\phi \to
-E\,\alpha\sin\thc$ and $k_\theta \to E\,\beta$, so $(\alpha,\beta)$ reduce to the standard celestial coordinates $(\alpha_{\rm B},\beta_{\rm B})$ of
Bardeen~\cite{Bardeen:1973}, $\alpha_{\rm B}\equiv -k_\phi/(E\sin\thc)$ and
$\beta_{\rm B}\equiv k_\theta/E$. At finite $\rc$, the two differ at relative order
$M/\rc$.

To be more precise, for Kerr in Boyer--Lindquist coordinates, the exact relation between our screen coordinates $(\alpha,\beta)$ and Bardeen's $(\alpha_{\rm B},\beta_{\rm B})$ is:
\begin{equation}
  \alpha_{\rm B} = \frac{\alpha\,\sqrt{\Delta_{\rm c}}/\mathcal{D} + 2Ma\,\rc\sin\thc/\Sigma_{\rm c}}{1-2M\rc/\Sigma_{\rm c}}\,,
  \qquad
  \beta_{\rm B} = \frac{\beta\,\sqrt{\Sigma_{\rm c}}}{\mathcal{D}\,\sqrt{1-2M\rc/\Sigma_{\rm c}}}\,,
  \label{eq:FOORTtoBardeen}
\end{equation}
where $\mathcal{D}=\sqrt{\rc^2+\alpha^2+\beta^2}$, and $\Sigma_{\rm c}$ and $\Delta_{\rm c}$ are the relevant Kerr metric functions, evaluated at the camera coordinates $(\rc,\thc)$. For large $\rc$, this becomes
\begin{equation}\label{eq:coordstoBardeencoords}
  \alpha_{\rm B} = \alpha\left(1+\frac{M}{\rc}\right) + \frac{2aM\sin\thc}{\rc} + \mathcal{O}(\rc^{-2})\,,
  \qquad
  \beta_{\rm B} = \beta\left(1+\frac{M}{\rc}\right) + \mathcal{O}(\rc^{-2})\,.
\end{equation}
The isotropic factor $(1+M/\rc)$ is the gravitational blueshift $1/\sqrt{-g_{tt}}$ at the
camera. The $a$-dependent shift in $\alpha$ is a frame-dragging effect since the camera has $e_t\propto\partial_t$ and thus is a static observer instead of a zero-angular-momentum observer. Note that a consequence of Eq.~\eqref{eq:coordstoBardeencoords} is that screen sizes measured in {\tt FOORT}'s screen coordinates, such as ring diameters
compared with analytic critical curves, are smaller by a factor $1-M/\rc$. For a  camera distance $\rc=1000M$ ({\tt FOORT}'s default), this is typically sub-pixel level for reasonable image resolutions --- e.g.\ this is undetectable in Fig.~\ref{fig: example nb shadow}.

\subsection{Output \& analysis}\label{sec: output and analysis}

The geodesic integrator will keep on computing trajectories until predetermined \textit{termination conditions} are met. Depending on the metric of interest, different terminations for the geodesic integrator can be included:
\begin{itemize}
    \item \textbf{Boundary sphere:} The geodesic integrator will stop once the geodesic arrives at a (user-defined) boundary sphere with a tunable radius. This should be large enough such that geodesics reaching this boundary sphere can really be considered as ``escaped to infinity''.
    \item \textbf{Time out:} One can set a limit on the number of integration steps that are taken per geodesic. This is particularly useful for geodesics that approach bound orbits or coordinate singularities.
    \item \textbf{Event horizon:}  For certain metrics with an event horizon (currently only horizons at a fixed value of the radial coordinate), a parameter $\epsilon_H$ can be set such that the geodesic integration is terminated once the geodesic reaches $r_H(1+\epsilon_H)$.
    \item \textbf{$\theta$-singularity:} Metrics where one of the coordinates is the polar angle $\theta$ typically become singular at the poles. To avoid numerical instability, geodesic integration is terminated once the geodesic position approaches the poles within a tunable $\Delta\theta < \epsilon_\theta$.
    \item \textbf{General singularity:} Metrics can contain any number of (naked) singularities, of any codimension. A parameter $\epsilon_S$ can be set such that geodesic integration will  be terminated when geodesics come within a flat distance of a singularity.\footnote{More precisely, with the singularity given by $x^{\mu_i}=x_s^{\mu_i}$ for $i=1,...$ (with the range of $i$ depending on the singularity's codimension), the geodesic at position $x_g^\mu$ will be terminated if $\sum_i|x_g^{\mu_i}-x_s^{\mu_i}|^2\leq\epsilon_S^2$.}
\end{itemize}
Additionally, geodesic integration can also be terminated when a \texttt{NaN} is encountered.\\

\begin{figure}
    \centering
    \includegraphics[width=\linewidth]{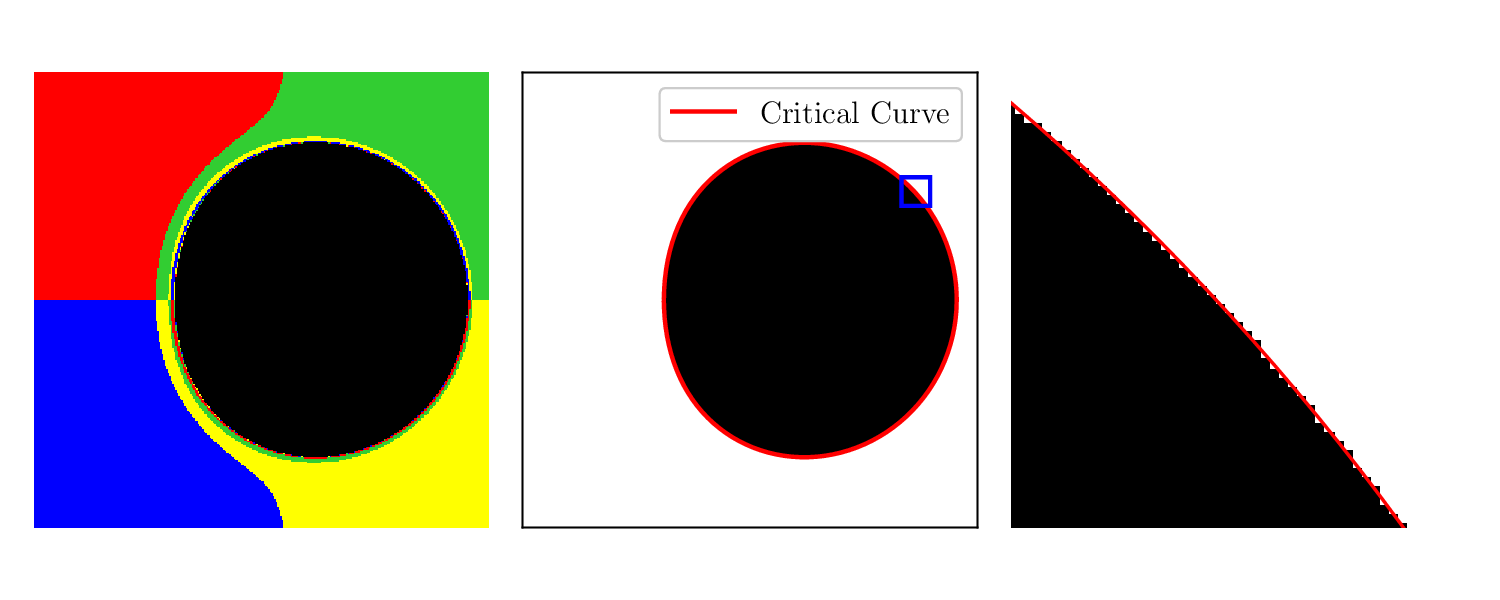}
    \caption{Comparison of the BH shadow, as extracted by \texttt{FOORT}, with the analytic prediction for a Kerr BH with spin $a = 0.9M$. From left to right, we show the four-color screen, the extracted shadow and predicted critical curve, and a zoomed-in version focussing on the blue square. The {\tt FOORT} data uses RK4 integration with $\epsilon = 0.03$, and the critical curve is corrected for the finite distance through Eq.~\eqref{eq:coordstoBardeencoords}.}
    \label{fig: example nb shadow}
\end{figure}

\noindent At the moment, \FOORT allows the user to track the following \textit{diagnostics}:
\begin{itemize}
    \item \textbf{Geodesic position:} When switched on, \FOORT will keep track of the entire geodesic trajectory and output it to a file. This is useful e.g.\ to visualize individual geodesic trajectories.
    \item \textbf{Four-color screen:} This diagnostic assigns a number 1--4 to the geodesic based on the quadrant of the boundary sphere where the geodesic terminates (see left panel in Fig.~\ref{fig: example nb shadow}). If the geodesic does not reach the boundary sphere (e.g.\ due to terminating at a horizon or singularity), it will be assigned 0.
    \item \textbf{Equatorial passes:} This allows the user to track the number of times a geodesic passes through the equatorial plane $\theta = \pi/2$. If the geodesic escapes to infinity, this number will be non-negative: if it falls into a horizon, it will get a minus sign. This is typically useful in the context of lensing bands or photon rings, which are indexed by the number of times their light rays pass through the equatorial plane. 
    \item \textbf{Closest radius:} This diagnostic keeps track of the smallest value for the radial coordinate a geodesic encounters. If the geodesic falls into a horizon, the code will report $r=0$.
    \item \textbf{Equatorial emission:} This keeps track of the brightness intensity of the geodesic assuming a particular model for emission from an equatorial disk. The calculation of this brightness intensity is described in more detail in Sec.~\ref{sec: emission}.
\end{itemize}

These different terminations and diagnostics allow the user to study the geometry of exotic spacetimes, and enable the analysis of their potential photon rings even though this may not be feasible analytically. 
The full documentation provides details on how to use these different diagnostics and terminations, as well as how to add new diagnostics and terminations to {\tt FOORT} in a few simple steps.

The {\tt FOORT} repository contains a Python module, {\tt pyFOORT}, which provides helper functions to easily visualize raw output of these diagnostics with standard Python packages. Since the output of an adaptive mesh is not on a regular pixel grid, {\tt pyFOORT} first interpolates the raw data onto a uniform grid; from this, a provided function for every image-type output listed above can convert the raw data file to an image. {\tt pyFOORTInterferometry} is another included Python module which can output visibility amplitudes based on {\tt FOORT} data (see Sec.~\ref{sec: emission} below). Specific parameters and options for the use of all diagnostics and terminations, as well as {\tt pyFOORT} and {\tt pyFOORTInterferometry}, are described in the extended documentation on the repository.

\subsection{Adaptive mesh integration}

\FOORT features a specific adaptive mesh which allows  high image resolutions to be achieved by efficiently choosing the geodesics to integrate. Our adaptive mesh algorithm starts by integrating an evenly spaced, square grid of pixels. Each pixel is then assigned a ``weight'', based on how much the specified image feature changes in its nearby pixels. Based on all of these pixel weights, at each iteration the algorithm decides to divide the squares associated with pixels of the highest weights.

The tuneable parameters of the algorithm are the initial number of pixels to integrate, $\nin^2$; the maximum number of subdivisions that may be made, $\smax$ (the initial grid is considered to be the first subdivision, so necessarily $\smax\geq 1$), the maximum number of pixels allowed to integrate $\nmax$ (with necessarily $\nin^2\leq \nmax$); and the number of pixels to subdivide each iteration, $n_\mathrm{iter}$.

\begin{figure}[htbp]\centering
 \includegraphics[width=0.4\textwidth]{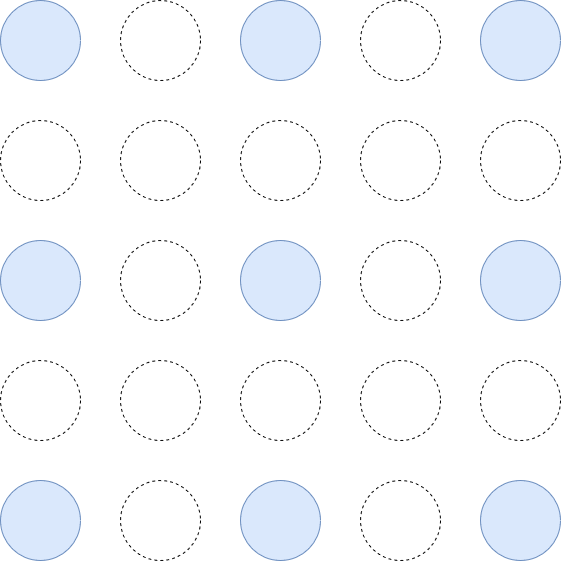}
 \quad \quad \quad
 \includegraphics[width=0.4\textwidth]{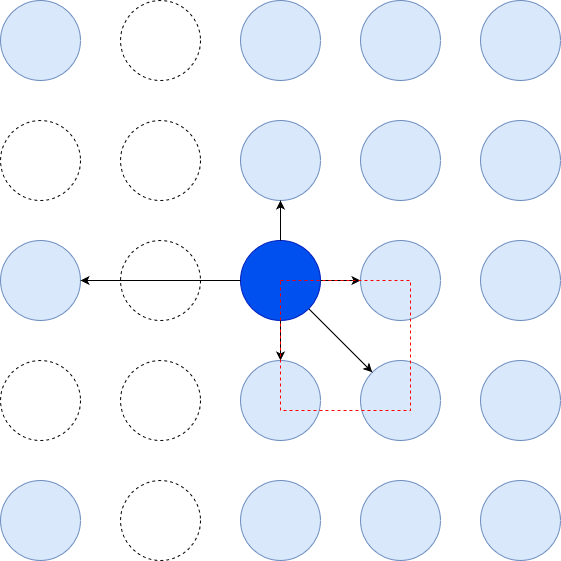}
 \caption{(\textit{left}) The total grid generated for $\nin = 3, \smax = 2$ has 25 pixels total. The initial pixels integrated in Step 0 are indicated in blue. (\textit{right}) The grid with $\nin = 3, \smax = 2$ after some number of integration iterations. The integrated pixels are indicated in blue. The neighbors of the dark blue central pixel are given by the arrows, and its ``square'' is given by the four pixels in the red dashed lines. Note that the dark blue pixel has $s=\smax=2$.
 }
 \label{fig:mesh}
\end{figure}

The algorithm can then be broken down into steps as follows:
\begin{itemize}
 \item[\textbf{Step 0.}] Integrate a $\nin\times \nin$ square grid of pixels.
 
 Note that the mesh algorithm creates a grid that is already the maximum possible size,  i.e. $(\nin-1)*2^{\smax-1} +1$. Pixel row and column coordinates then lie between $0$ and $(\nin-1)*2^{\smax-1}$; the initial grid is all pixels with coordinates $(i,j)\times 2^{\smax-1}$ for $0\leq i,j\leq \nin-1$; see Fig.~\ref{fig:mesh}.

 \item[\textbf{Step 1.}] Calculate the ``weight'' of all pixels, i.e.\ a quantitative measure of how much a specified image feature changes in nearby pixels.

A particular diagnostic - i.e.\ those detailed in Sec.~\ref{sec: output and analysis} -  is set as the mesh diagnostic in the \FOORT configuration. This mesh diagnostic assigns a \emph{value} to each geodesic at the end of its integration; this value can be a vector of real numbers of any size (determined by the particular diagnostic). Every diagnostic also specifies a way of determining a \emph{distance} between such values. The \emph{weight} of a pixel is then determined as the maximum of the \emph{distances} between itself and its lower, right, and low-right neighbors --- i.e.\ between itself and the other three pixels that define its ``square'' (see Fig.~\ref{fig:mesh}).

For example, the diagnostic \code{FourColorScreen} assigns a number between 1 and 4 to each geodesic as its \emph{value}, depending on which color has been assigned. The distance between two pixels is then the discrete metric, i.e.\ 0 if their colors are the same and 1 if they are different. Another example is the diagnostic \code{GeodesicPosition}: this assigns to a pixel as value the angular coordinates $(\theta,\phi)$ of the position of the geodesic at termination. The distance between two pixels is then the great-circle angle on the unit sphere between these two sets of coordinates.

 \item[\textbf{Step 2.}] Order all pixels by their weight and select the $n_\mathrm{iter}$ highest-weight pixels.
 \item[\textbf{Step 3.}] Subdivide the squares associated with all of these highest-weight pixels and integrate geodesics for these pixels.
 \item[\textbf{Step 4.}] Go back to step 1 as long as the total number of geodesics integrated is still smaller than $\nmax$.
\end{itemize}

\FOORT keeps track of all pixels and their properties in a specific \code{struct}. Every pixel keeps track of:
\begin{itemize}
 \item Its position on the screen (i.e.\ pixel coordinates);
 \item Its weight;
 \item Its subdivision level $s$ ($s=1$ for the initial grid and $s=\smax$ for the finest level). A pixel at level $s$ has its right and lower neighbors at a distance of $2^{\smax-s}$. The pixels on the rightmost column and bottom row are assigned $s=\smax$; they have no right or lower neighbors.
 \item A pointer to the pixel \code{struct} of its right, lower, and lower-right neighbors; these are always at a distance governed by the pixel's subdivision $s$. Together with itself, these four pixels define the pixel's own ``square''; see Fig.~\ref{fig:mesh}.
 \item A pointer to the pixel \code{struct} of its upper and left neighbor; these are solely defined as being \emph{reciprocal} with respect to the right- and lower-neighbors. For example, pixel $p_1$ has $p_2$ as its \emph{left} neighbor if $p_2$ has $p_1$ as its \emph{right} neighbor; see Fig.~\ref{fig:mesh}.
\end{itemize}
Having each pixel \code{struct} contain pointers to its neighbor \code{structs} allows efficient traversal of the grid around any given pixel when needed.

\subsection{Emission models and interferometry}\label{sec: emission}

\FOORT has the option to model an accretion disk in the spacetime of interest, serving as a light source whose photons are ray-traced to the observer. At the moment, the only emission model that is implemented is the \texttt{GLMJohnsonSU} model \cite{gralla_shape_2020}. It describes a geometrically thin disk in the equatorial plane, with an intensity profile given by
\begin{equation}\label{eq: GLM model}
    I_0(r) = \left((r-\mu)^2+\sigma^2\right)^{-1/2}  \exp\left(-\frac12 \left[\gamma + \operatorname{arcsinh}\left(\frac{r-\mu}{\sigma}\right)\right]^2\right)\,,
\end{equation}
where the parameters $\mu$ and $\sigma$ control the peak and the width of the profile, whereas $\gamma$ controls the skewness of the profile: a negative value makes the profile fall off rapidly at smaller radii, while maintaining a more gradual slope outside the peak radius. While this is a highly simplified, overly symmetric model for emission from an accretion disk, it has nevertheless proven useful to understand certain features of BH images and especially the BH photon rings \cite{johnson2020universal,gralla_shape_2020,Cardenas-Avendano:2022csp,Lupsasca:2024wkp}.\footnote{Note that the emission profile is determined using the $r$ coordinate in Eq.~\eqref{eq: GLM model} for any metric; therefore, care must be taken when comparing emissions across different metrics, especially if these have different masses.}

The total brightness intensity for the geodesic is calculated as
\begin{equation}\label{eq: I tot}
    I_{\text{tot}} = \sum_{n=1}^{N_{\text{max}}} \zeta_n I_0(p^{(n)}) g(p^{(n)})^f\,.
\end{equation}
In this expression, $N_{\text{max}}$ is the number of times the geodesic passes through the equatorial plane\footnote{$N_{\text{max}}$ can also be set to a fixed value, thus suppressing higher-order rings.} and $\zeta_n$ is the geometric fudge factor used to mimic the effects of a geometrically thick disk: $\zeta_1 = 1$ and $\zeta_{n>1} = \zeta$.
$I_0(p^{(n)})$ is the local intensity at the $n$th point of passing $p^{(n)}$; $g(p^{(n)})$ is the redshift factor from the relative motion between the accretion disk fluid and an observer at infinity. The power $f$ of the redshift is set to 3 by default\footnote{A power of 3 is a common choice in the literature, as $I_\nu / \nu^3$ is the invariant phase-space occupation number of photons of frequency $\nu$ along a ray. With $f=3$, $I_{\text{tot}}$ is to be interpreted as a specific intensity under the assumption of a flat spectrum ($I_\nu\sim \nu^0$) source. 
Setting $f=4$ means that $I_{\text{tot}}$ is to be interpreted as bolometric \cite{gralla_shape_2020}.}. 

In order to determine the redshift factor $g(p^{(n)})$, we need to assume a model for the velocity of the emitting matter. The \texttt{GeneralCircularRadialFluid} model in \FOORT allows for a mix of (sub)Keplerian circular orbits and radially infalling orbits. 
It takes three parameters: the sub-Keplerian parameter $\xi$ and the mixing parameters $\beta_r$ and $\beta_\phi$, all in $[0,1]$. The model is briefly summarized below, following Refs.~\cite{Cardenas-Avendano:2022csp, Pu:2016flow}, and detailed further in App.~\ref{app: fluid velocity}.

If the ISCO exists, it is first found using a bisection method. The conserved quantities associated with this ISCO are determined and saved. Then, to determine the equatorial fluid four-velocity $w(p)$ at a given point $p$ (i.e., a point of passage of the geodesic through the equatorial plane), we expand it as
\begin{align}
    w & = w^t \partial_t + w^r\partial_r+w^\phi\partial_\phi\\
    & \equiv w^t \left(\partial_t - \iota \partial_r + \Omega \partial_\phi\right),
\end{align}
and proceed to find $w^t,\iota,\Omega$ using the following procedure:
\begin{enumerate}
    \item First, we determine the ``circular velocity'' vector $w_{\text{cir}}$ at the point $p$. The calculation depends on whether the radial coordinate of the point $p$ is in- or outside the ISCO, as detailed in App.~\ref{app: fluid velocity}, and involves optional rescaling with the sub-Keplerian parameter $\xi$.
    \item The radial infall velocity vector $w_{\text{rad}}$ is calculated at $p$.
    \item The final velocity vector is now calculated as a mix of the two above using the mixing parameters $\beta^r, \beta^\phi$ as
    \begin{align}
        w^r & = w_{\text{cir}}^r + (1- \beta_r)(w_{\text{rad}}^r-w_{\text{cir}}^r) \,, \label{eq: uu_r general} \\
        \Omega & = \Omega_{\text{cir}} + (1- \beta_\phi)(\Omega_{\text{rad}}-\Omega_{\text{cir}}) \,,  \label{eq: Omega general}\\
        w^t & = \sqrt{\frac{-1 - g_{rr} (w^r)^2}{g_{tt} + 2 g_{t\phi} \Omega + g_{\phi\phi}\Omega^2}}\,, \label{eq: fix uu_t} 
    \end{align}
\end{enumerate}
where \eqref{eq: fix uu_t} follows from the normalization $w_\mu w^\mu = -1$. 
It should be noted that, in general, the flow described by Eqs.~\eqref{eq: uu_r general}--\eqref{eq: fix uu_t} is not geodesic.
With this model for the fluid velocity, the redshift factor can now be determined as
\begin{equation}
    g(p^{(n)}) = \frac{-E}{(-u^\mu)\, w_\mu(p) }\,,
\end{equation}
where $u$ represents the current velocity of the geodesic; the extra minus sign in the equation is necessary since $u^\mu$ is past-pointing (so $-u^\mu$ is future-pointing).

\medskip

The resulting intensity profile on the observer screen can be transformed to Fourier space in order to obtain the visibility --- the quantity actually measured in VLBI experiments. The complex visibility is defined as
\begin{equation}
    V(\boldsymbol{u}) = \int I(\boldsymbol{x})\, e^{-2\pi i\, \boldsymbol{x}\cdot \boldsymbol{u}} \: \dd^2 \boldsymbol{x}\,,
\end{equation}
where $\boldsymbol{x}$ denotes angular coordinates on the sky (in radians), measured from the center of the observer screen, and $\boldsymbol{u}$ is the baseline projected onto the sky plane, in units of the observing wavelength. \FOORT comes with a Python module, \texttt{pyFOORTInterferometry}, that contains functions to convert \FOORT data into (normalized) visibility amplitudes $|V(\boldsymbol{u})|/V(0)$. In practice, the projection-slice theorem is used to compute the visibility along a line $L_\phi$ in the Fourier plane that passes through the origin ($|\boldsymbol{u}| = 0$) at an angle $\phi$ measured counterclockwise with respect to the horizontal axis. To this end, the image is first projected onto the line through the image-plane origin at the same angle $\phi$, i.e.\ its Radon transform is computed by integrating the intensity along lines perpendicular to that direction; this one-dimensional projection is then Fourier transformed.

The visibility can then be analyzed for universal signatures of the underlying spacetime, such as properties of the photon rings \cite{johnson2020universal,Lupsasca:2024wkp}; in particular, the shape and slope of the visibility amplitude are determined by the photon rings and their properties. The dependence of these signatures on the compact-object spacetime can be efficiently explored and compared with \FOORT without the need for expensive GRMHD simulations; of course, the visibility functions obtained with \FOORT in this way, using a simplified emission model, do not reflect the complexity of realistic accretion disks.

\section{Demonstrations \& Applications}\label{sec: demos}

First of all, we verify that {\tt FOORT} can reproduce an analytic BH shadow. As an example, Fig.~\ref{fig: example nb shadow} compares the shadow, as extracted by \texttt{FOORT}, to the analytic prediction (see App.~\ref{app: metrics > BHs > Kerr} for details) for a Kerr BH with spin $a=0.9M$. The zoomed-in panel demonstrates that {\tt FOORT} reproduces the expected curve to the pixel level.

\begin{figure}
    \centering
    \includegraphics[width= 0.9\linewidth]{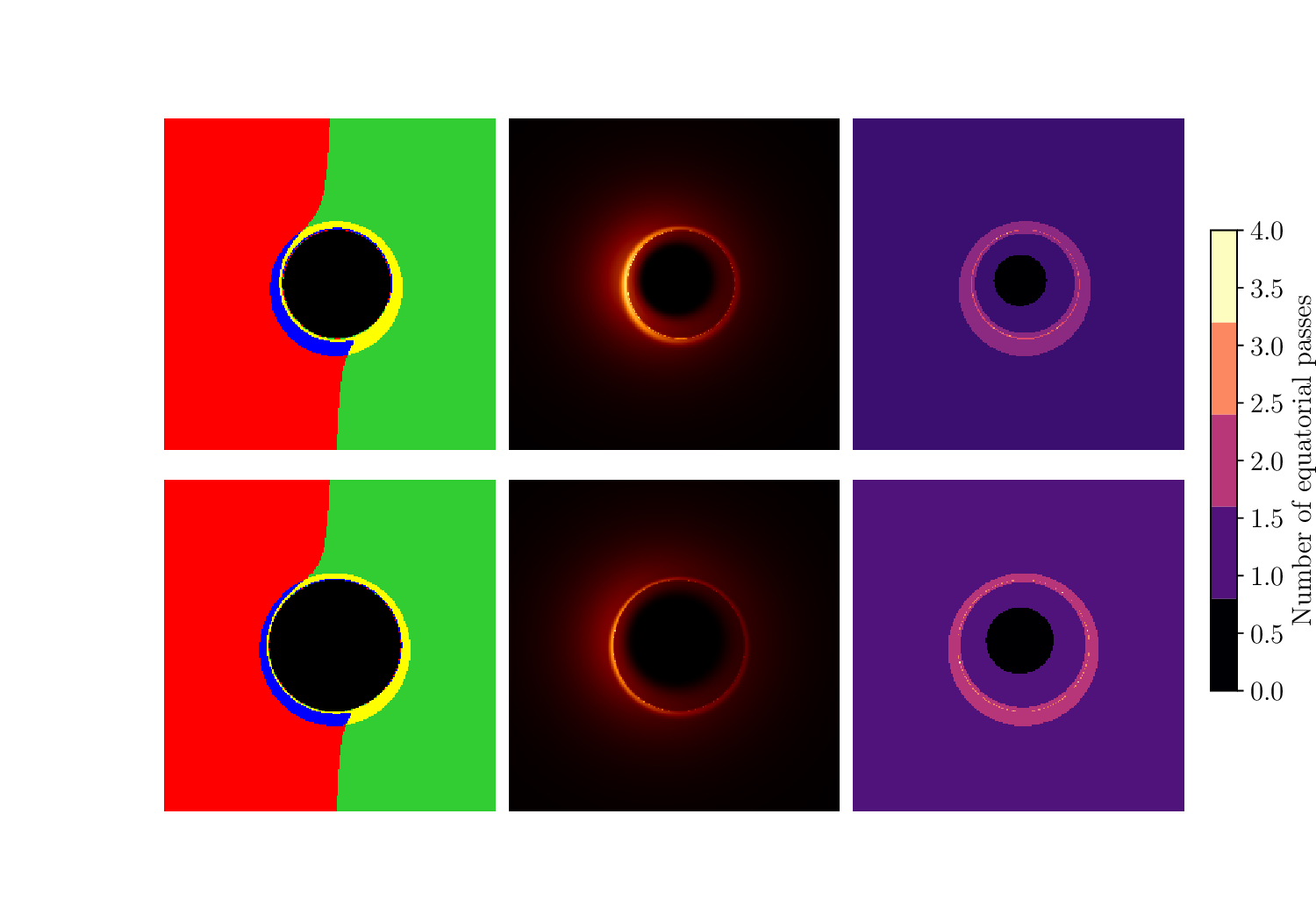}
    \vspace{-1cm}
    \caption{From left to right, we show the four-color screen, the emission and the number of equatorial passes for (\textit{top}) a Kerr BH with spin $a=0.94M$ and (\textit{bottom}) the Johannsen BH with the same spin and deviation parameter $\alpha_{13} = 5.5$, as seen by an observer at inclination $\theta_0 = 17^\circ$. The screen size is $30M \times 30M$. These examples have moderate resolution, and are provided in the repository.}
    \label{fig: example nb images}
\end{figure}

We continue by visualizing the Kerr spacetime with spin $a=0.94M$, as seen by an observer at inclination $\theta_0 = 17^\circ$. We use the GLM model \eqref{eq: GLM model} with $\mu = 2.02M$ (the radial coordinate of the ISCO), $\sigma = 0.5M$, and $\gamma=-2$. The fluid is assumed to be Keplerian ($\xi = \beta_r = \beta_\phi = 1$).
The top row in Fig.~\ref{fig: example nb images} shows --- from left to right --- the four-color screen, the emission, and the number of equatorial passes.
The emission contains the expected photon ring, which can be linked to the geodesics that pass through the equatorial plane multiple times, as shown by the rightmost column.

\medskip

As an example, we also provide the same images for the Johannsen metric~\eqref{eq: Johannsen metric} with deviation parameter $\alpha_{13}=5.5$; as the latter is rather large, the changes are clearly visible. Note that we kept the value of $\mu$ fixed to the Kerr ISCO, whereas the ISCO of this Johannsen BH is now $r_{\text{ISCO}} \approx 5.54M$; this is automatically taken into account in {\tt FOORT}. 

\begin{figure}[ht]
    \centering
    \includegraphics[width=0.8\linewidth]{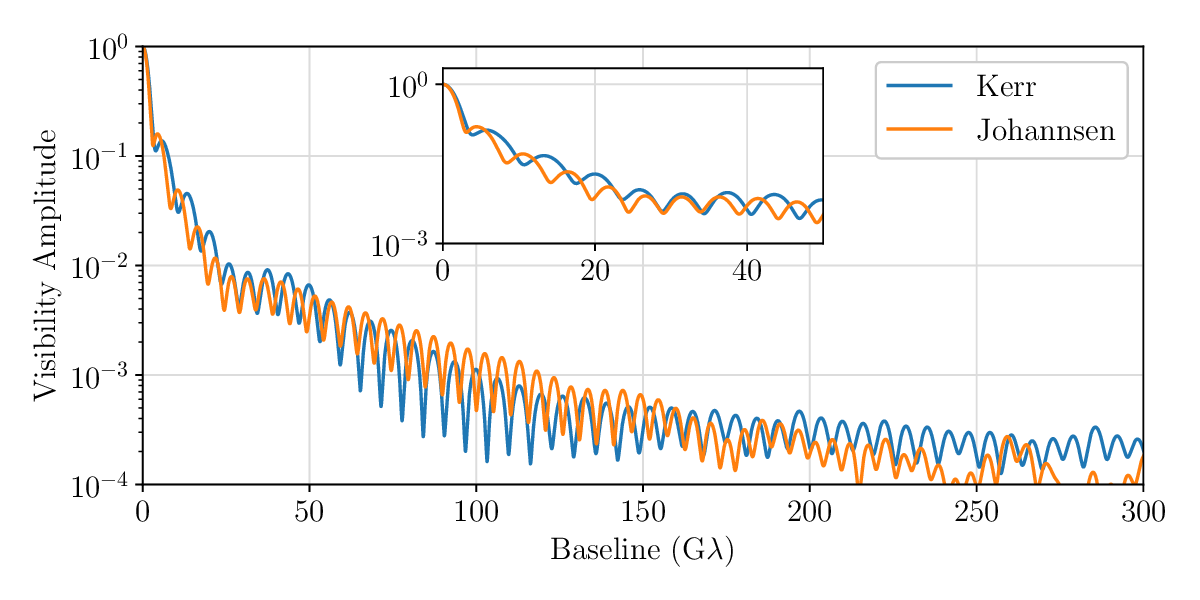}
    \caption{Normalized visibility amplitudes associated with the emission images in Fig.~\ref{fig: example nb images}, along an angle of $40^\circ$. The images were assumed to have a width of 100 $\mu$as, i.e.\ an angular size of $\theta_g = GM/(c^2D) = 3.33\mu$as.}
    \label{fig: example nb visamps}
\end{figure}

Figure~\ref{fig: example nb visamps} shows visibility amplitudes associated with both emission images in Fig.~\ref{fig: example nb images}, extracted using the built-in functions in the \FOORT repository. Given that the images were already visibly distinct, it is no surprise that the visibility amplitudes start deviating early on too. To obtain highly accurate results (at large baselines), these visibility amplitudes should be obtained from high-resolution versions of the images in Fig.~\ref{fig: example nb images}.\footnote{Additionally, in the $30M$ screen used here, about $10\%$ of the disk flux falls outside of the image, which biases the normalized visibility amplitudes upward. A higher field of view could be used to improve this normalization as well.}

\medskip

\begin{figure}[ht]
    \centering
    \includegraphics[width=0.4\linewidth]{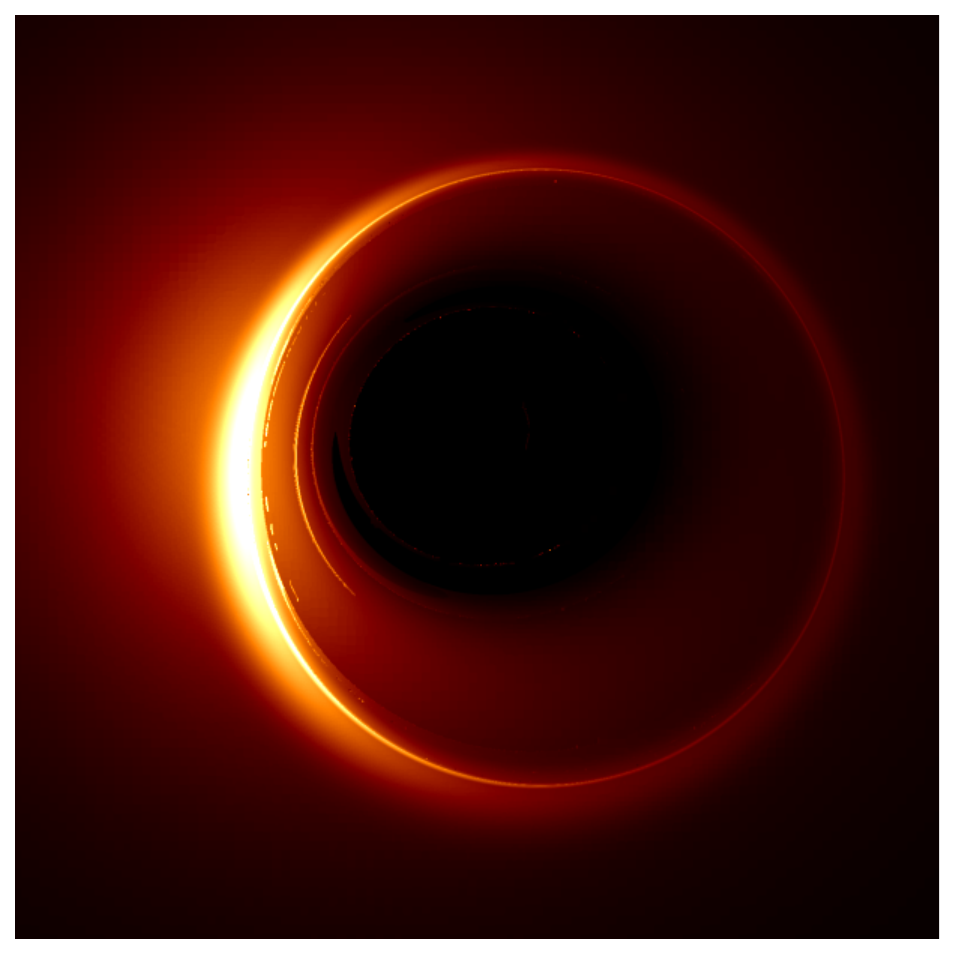} \includegraphics[width=0.4\linewidth]{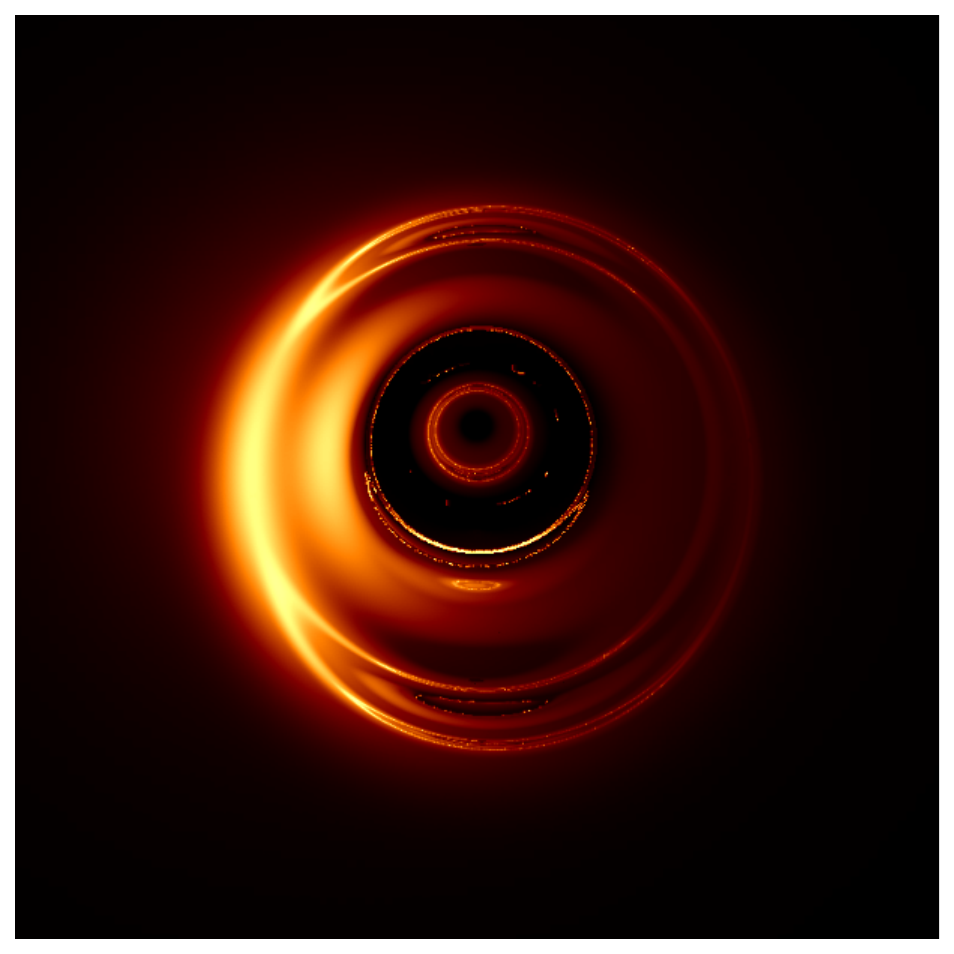}
    \caption{Images of a Manko--Novikov BH (left) and a ring fuzzball (right). The Manko--Novikov BH has parameters $a/M=0.94$ and $\alpha_3=2$; see App.~\ref{app:MN}. The ring fuzzball has parameters $P=2,q_0=50,\lambda=0.19$; see App.~\ref{app:FB}. Both are imaged at inclination $\theta_0=30^\circ$ and with emission parameters $\xi=\beta_r=\beta_\phi=1$, $\gamma=-1.5$, and with $\mu=0.66$, $\sigma=0.5$ (for Manko--Novikov) or $\mu=-0.5$, $\sigma=2.0$ (for the fuzzball).
     Reproduced from Ref.~\cite{mayerson2023observational}.}
    \label{fig: MN FB}
\end{figure}

{\tt FOORT} is specifically constructed to allow for ray-tracing more exotic metrics for objects that can either have horizons, singularities, or both. As an example, in Fig.~\ref{fig: MN FB}, we show ray-traced images of a Manko--Novikov BH (see App.~\ref{app:MN}) and a so-called ``ring fuzzball'' (see App.~\ref{app:FB}). The Manko--Novikov BH can be seen as a multipole-deformed vacuum black hole and has a horizon which is partly singular. The fuzzball metric is a horizonless BH mimicker which is constructed in string theory as a smooth metric in higher dimensions; however, it consists of naked singularities in four dimensions.

\begin{figure}[ht]
    \centering
    \includegraphics[width=0.8\linewidth]{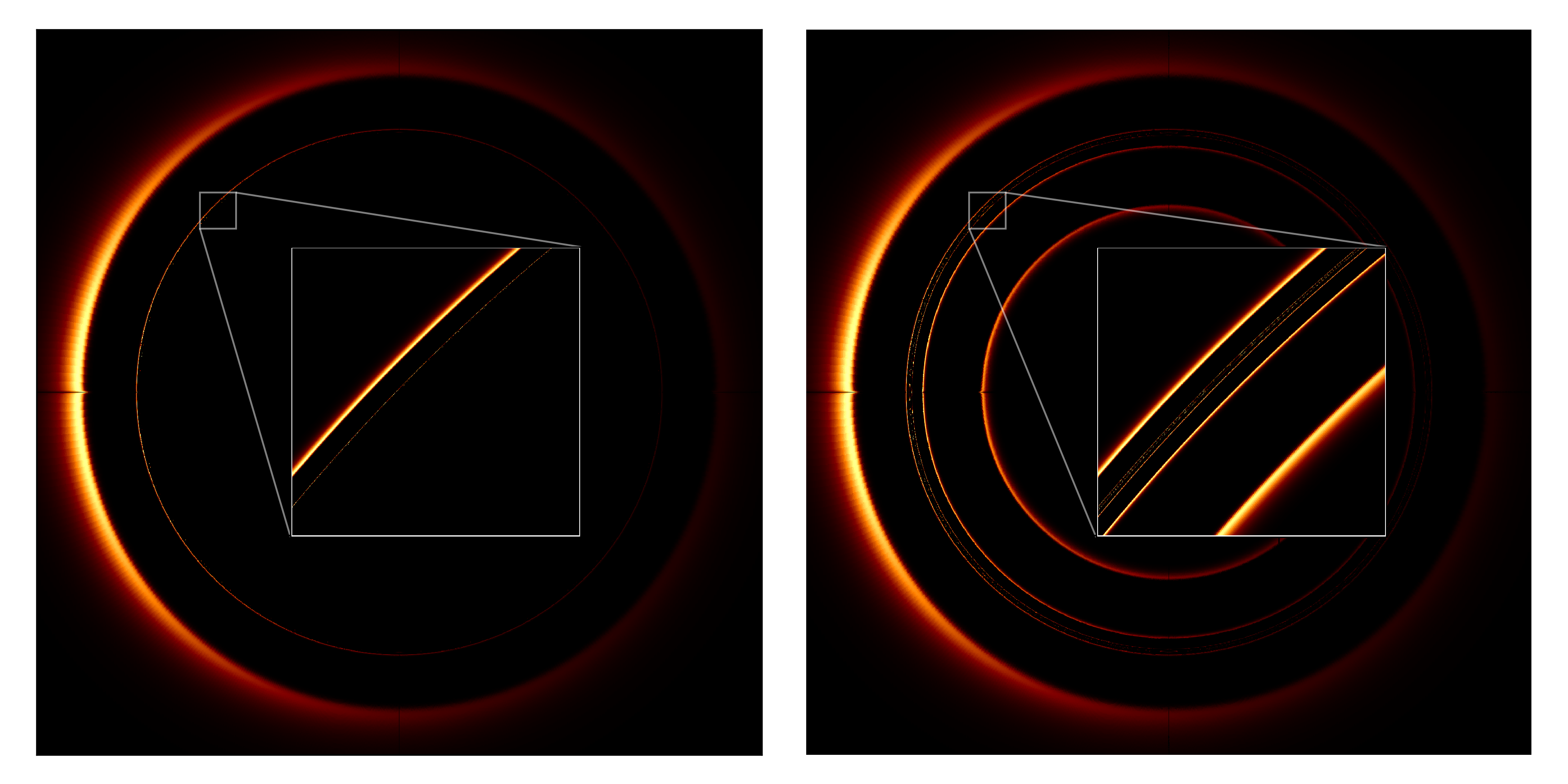}
    \caption{Side-by-side comparison of ray-traced images of a (\textit{left}) Schwarzschild BH and (\textit{right)}
    an ultracompact solitonic BS of the same ADM mass. Both panels also provide a zoom-in of the
    photon ring region, exhibiting their fractal structure. Reproduced from Ref.~\cite{Marks:2025jpt}.}
    \label{fig: BH BS comp}
\end{figure}

Aside from analytical metrics, {\tt FOORT} can also be used to ray-trace numerically obtained metrics, as long as they can be read in from files. An example is the solitonic boson star (BS) metric (see App.~\ref{sec: sol BS}), which is sufficiently compact to have a pair of light rings. Figure~\ref{fig: BH BS comp}, reproduced from Ref.~\cite{Marks:2025jpt}, shows a comparison of this BS and a Schwarzschild BH of the same mass, where the emission is provided by an equatorial disk described by Eq.~\eqref{eq: GLM model}. The images are obtained with the Verlet scheme, for an observer located at areal radius $\mu_{\rm BS} r = 1000$ in units where the mass of the boson star is $\mu_{\rm BS} M \approx 10.42$. The image width in these units is $150\times150$ in the same units, and uses 5 refinement levels. The insets focus on a region of $7.5\times7.5$, with 4 refinement levels. The disk parameters are fixed to $(\mu, \sigma, \gamma) = (60, 2.5, -2)$, with the first value corresponding to $\sim 6M$, the radial coordinate of the ISCO in Schwarzschild spacetime: the ultracompact BS has a \textit{marginally stable circular orbit} (MSCO) here, but it is not the ISCO, as circular orbits with $r \lesssim 3M$ are also stable. The fluid parameters are $\xi = \beta_r = \beta_\phi = 1.0$, corresponding to pure Keplerian orbits. Cunningham's prescription (see App.~\ref{app: fluid velocity}) is still applied from the MSCO inwards, as the intensity model \eqref{eq: GLM model} with the above parameters is approximately zero at these inner stable circular orbits. Due to the ultracompact nature of the BS, having a light ring at approximately the Schwarzschild value $r = 3M$, the photon ring structure outside the critical curve matches that of the BH very well. Inside the critical curve, the absence of a horizon creates additional photon rings: this gives rise to low-frequency modulations in the visibility amplitude --- see Ref.~\cite{Staelens:2026rzb} for more details.

\section{Conclusion}\label{sec: conclusion}
This paper presents {\tt FOORT}, a highly flexible and modular ray-tracing code that is specifically designed to visualize spacetimes beyond the standard Kerr black hole. We have provided details on some of the numerical techniques {\tt FOORT} implements, and demonstrated good scaling and convergence properties through various tests. Furthermore, we presented in detail its adaptive mesh algorithm, which can provide high resolution in regimes where any desired diagnostic changes sufficiently. Finally, we discussed the implementation of the GLM intensity model, as well as the generalized Keplerian fluid that we assume to source the emission.

We have showcased how \FOORT can be used to image a variety of spacetimes, ranging from modified black holes to boson stars and fuzzballs, and how the accompanying Python packages can be used to do photon-ring interferometry.

There are numerous ways {\tt FOORT} can be upgraded --- which the authors explicitly encourage, in line with {\tt FOORT}'s philosophy of flexibility and modularity. Emission models can be extended from {\tt FOORT}'s current implementation of equatorial, geometrically thin disks, to handle e.g.\ volumetric radiative transfer or time-dependent sources. {\tt FOORT} could also relatively easily be modified to keep track of photon polarization, or add source terms to the geodesic equation. More ambitious expansions could include implementing {\tt FOORT} within frameworks exploiting parallelization beyond CPUs and the use of OpenMP.

\section*{Code availability}
\FOORT (version {\tt 1.0.0})  \cite{Mayerson_FOORT_Flexible_Object-Oriented_2026} is publicly available at \url{https://github.com/drmayerson/FOORT} and free to use under the MIT license. Further documentation is provided in the repository.

%%%%%%%%%%%%%%%%%%%%%%%%%%%%%%%%%%%%%%%%%%%%%%
%%%%%%%%%%%%%%%%%%%%%%%%%%%%%%%%%%%%%%%%%%%%%%
%%%%%%%%%%%%%%%%%%%%%%%%%%%%%%%%%%%%%%%%%%%%%%
\section*{Acknowledgments}
We especially want to thank everyone who has tested \FOORT in some capacity or contributed to various features: A.\ C\'ardenas-Avenda\~{n}o (for assistance in visibility amplitudes and the Python implementation of their generation), L.\ Van Dael (for implementing the fuzzball metric), P.\ Fernandes (for the Apple makefile, contributions to the Python plotting scripts, and testing/debugging various features related especially to Kerr-Schild coordinates), S.\ Giri (for the Arch makefile and testing/debugging various features).

\medskip

Additionally, we would like to thank J.\ Davelaar, S.\ Gralla, T.\ Hertog, L.\ K\"uchler, T.\ Lemmens, T.\ Li, A.\ Lupsasca, O.\ McGloin, H.\ Olivares, H.\ Paugnat, B.\ Ripperda, B.\ Vercnocke for interesting discussions and earlier ray-tracing collaborations.
D.R.M.\ acknowledges support from the IBOF grant IBOF/21/084, Leuven Gravity Institute grant TPU-E6297-Incubatiefinanciering Inst LGI, and FWO grant QPG-3E4102-HERTOG-FWO-PROJECT TEAM ET.
S.J.S.\ is supported by the Centre for Doctoral Training at the University of Cambridge funded through STFC (ST/W006812/1, 2882538).
F.B.\ acknowledges support from the FED-tWIN programme (profile Prf-2020-004, project ``ENERGY'') issued by BELSPO, and from the FWO Junior Research Project G020224N granted by the Research Foundation -- Flanders (FWO)..
This work is also partially supported by the KU Leuven C1 grant C16/25/010. Computing resources were provided and supported by the VSC (Flemish Supercomputer Center), funded by the Research Foundation Flanders (FWO) and the Flemish Government -- department EWI.

\appendix

\section{Implemented metrics}\label{app: metrics}

\FOORT comes with several implemented BH metrics, as well as a fuzzball and boson star example. The modularity of the code makes it straightforward to add new metrics, as explained in the documentation. This appendix summarizes the spacetime metrics that are currently available.

As with the other {\tt FOORT} configuration options, such as the diagnostics and terminations described in Sec.~\ref{sec: output and analysis}, a detailed description of how to configure these metrics, their parameters, and options in practice is described in the extended documentation on the repository.

\subsection{Flat space}

As a baseline, \texttt{FOORT} has the Minkowski (flat space) metric implemented in the form
\begin{equation}
    \dd s^2 = -\dd t^2 + \dd r^2 + r^2\left(\dd\theta^2 + \sin^2\theta\,\dd\phi^2\right)\,.
\end{equation}
One can also choose to integrate the logarithmic radius $\Lambda = \log r$.

\subsection{Black holes}

\FOORT features a collection of BH metrics that inherit from a parent \texttt{SphericalHorizonMetric} class, which requires the existence of an event horizon at a fixed value of the radial coordinate $r=r_H$.

\subsubsection{Kerr}\label{app: metrics > BHs > Kerr}

The Kerr metric is implemented in standard Boyer--Lindquist coordinates:
\begin{empheq}[box=\fbox]{align}
    \dd s^2 =\, & - \left(1 - \frac{2M r}{\Sigma}\right) \dd t^2 + \frac\Sigma\Delta \dd r^2 + \Sigma \,\dd \theta^2 + \frac{\left(r^2 + a^2\right)^2 - \Delta a^2 \sin^2\theta}{\Sigma} \sin^2 \theta \, \dd \phi^2 \nonumber \\
    & - \frac{4 M r a \sin^2 \theta}{\Sigma} \, \dd t\, \dd \phi \label{eq: Kerr metric}
\end{empheq}
with 
\begin{equation}
    \Sigma = r^2 +a^2\cos^2\theta\,, \quad \Delta = r^2 + a^2 - 2M r,
\end{equation}
and $a \in [-M, M]$ the rotational parameter.

For the Kerr metric, the analytic expressions describing the \textit{critical curve} are known: this is the separatrix on the observer's screen between geodesics that fall into the event horizon and those that escape back to infinity, when ray-traced backwards \cite{Bardeen:1973}. We briefly list the expressions relevant to Fig.~\ref{fig: example nb shadow} below --- for more details, we refer the reader to e.g.\ Ref.~\cite{Lupsasca:2024wkp}.

The critical curve is parametrized\footnote{For conciseness, we restrict to the non-extremal case.} by a range of Boyer--Lindquist radii $[r_-^\gamma, r_+^\gamma]$ given by
\begin{equation}
    r_\pm^\gamma = 2M\left[1 + \cos\left(\frac23 \cos^{-1}\left(\pm \frac{a}{M}\right)\right)\right]\,.
\end{equation}
The impact parameters on the screen, for an observer at angle $\theta_0$, are
\begin{align*}
    \alpha & = -\frac{\lambda}{\sin\theta_0}\,, \\
    \beta & = \pm\sqrt{\eta + a^2\cos^2\theta_0 - \lambda^2 \cot^2\theta_0}\,,
\end{align*}
given in terms of the conserved quantities associated with spherical geodesics at radii in the range $[r_-^\gamma, r_+^\gamma]$,
\begin{align}
    \lambda & = \frac{M(r^2-a^2) -r\Delta}{a(r-M)}\,,\\
    \eta & = \frac{4r^2 \Delta}{(r-M)^2} - (\lambda - a)^2\,.
\end{align}

The metric is also implemented in Kerr--Schild coordinates\footnote{These are the ingoing coordinates in which Kerr originally wrote the metric \cite{Visser:2007fj}. Usually the name Kerr--Schild coordinates is reserved for a Cartesian version of these coordinates; with a slight abuse of notation we will also call these Kerr--Schild coordinates in {\tt FOORT}.}  obtained from the standard metric~\eqref{eq: Kerr metric} through the coordinate transformations \cite{Papadopoulos1998coordinates}
\begin{equation}\label{eq: coord transfo to Kerr coord}
    \dd t = \dd v - \frac{r^2+a^2}{\Delta} \dd r\,, \quad \dd\phi = \dd \bar{\phi} - \frac{a}{\Delta} \dd r\,,
\end{equation}
resulting in  
\begin{empheq}[box=\fbox]{align}
    \dd s^2 =\, & - \left(1 - \frac{2M r}{\Sigma}\right) \dd v^2 + 2 \dd v\dd r + \Sigma \,\dd \theta^2 + \frac{\left(r^2 + a^2\right)^2 - \Delta a^2 \sin^2\theta}{\Sigma} \sin^2 \theta \, \dd \bar{\phi}^2 \nonumber \\
    & - \frac{4 M r a \sin^2 \theta}{\Sigma} \, \dd v\, \dd \bar{\phi} -2a\sin^2\theta \dd r \dd \bar{\phi}\,. \label{eq: Kerr metric 2}
\end{empheq}

\subsubsection{Johannsen}

The Johannsen metric is a stationary and axisymmetric generalization of the Kerr metric that keeps a Carter-like constant and thus separability of the geodesic equations \cite{johannsen2013regular}. We follow \cite{Staelens:2023photonrings} in only keeping the first non-zero correction in the deviation functions, giving as metric
\begin{align}
    \dd s^2 =\,& -\Tilde{\Sigma}\frac{\Delta - a^2 A_2(r)^2 \sin^2\theta}{N(r,\theta)} \, \dd t^2  + \frac{\Tilde{\Sigma}}{\Delta A_5(r)}\, \dd r^2 + \Tilde{\Sigma}\, \dd \theta^2\nonumber\\ 
    & - 2 a\Tilde{\Sigma}\sin^2\theta \frac{(r^2+a^2)A_1(r)A_2(r) - \Delta}{N(r,\theta)}\, \dd t \, \dd \phi \label{eq: Johannsen metric}\\
    & + \Tilde{\Sigma} \sin^2 \theta \frac{(r^2+a^2)^2A_1(r)^2 - a^2 \Delta \sin^2\theta}{N(r,\theta)}\, \dd \phi^2\,,
    \nonumber
\end{align}
where
\be
\begin{aligned}
    A_1(r) & = 1+\alpha_{13}\left(\frac{M}{r}\right)^3\,,  & & &\Tilde{\Sigma}  & = r^2 + a^2\cos^2\theta + \epsilon_3 \frac{M^3}{r} \label{eq: dev functions Joh}
    \,,\\
    A_5(r) & = 1+\alpha_{52}\left(\frac{M}{r}\right)^2\,, & & &
    \Delta &  = r^2 - 2Mr + a^2 \,, \\
    A_2(r) & = 1+\alpha_{22}\left(\frac{M}{r}\right)^2\,, & & &
    N(r,\theta) & = \left[\left(r^2+a^2\right)A_1(r) - a^2A_2(r)\sin^2\theta\right]^2 \,.
\end{aligned}
\ee
This new metric depends on four additional parameters with respect to Kerr: $\alpha_{13}, \alpha_{22}, \alpha_{52}$, and $\epsilon_3$.

The Johannsen metric is asymptotically flat and reduces to Kerr when all deviation parameters $\alpha_{13}, \alpha_{22}, \alpha_{52}, \epsilon_3$ vanish. The event horizon of this BH is still the same as that of the Kerr metric, $r_H = M + \sqrt{M^2 - a^2}$. Demanding regularity outside of the horizon restricts the parameters as
\begin{align}
    \alpha_{52} & > -\frac{r_H^2}{M^2}\,, & 
    \epsilon_3 & > -\frac{r_H^3}{M^3} \,, &
    \alpha_{13} & > -\frac{r_H^3}{M^3}, \label{eq: bound a13}
\end{align}
together with the requirement that
\begin{equation}\label{eq: det req Joh}
    \alpha_{13} \neq \frac{a^2r(r^2+\alpha_{22}M^2)\sin^2\theta - r^3(r^2+a^2)}{M^3(r^2+a^2)}\,
\end{equation}
everywhere outside the horizon. 

\subsubsection{Manko--Novikov}\label{app:MN}
The Manko--Novikov metric is constructed as a solution to the vacuum Einstein equations with arbitrary multipole moments \cite{manko1992novikov}; it can be thought of as deforming the Kerr BH with (multipole) ``bumps''. Uniqueness theorems imply the solution is singular; in particular, the event horizon is singular at the equator. The general form of the Manko--Novikov metric, in Boyer--Lindquist-type coordinates, is
\begin{equation}\label{eq: MaNo metric}
    \dd s^2 = -f\left(\dd t - \omega \dd \phi\right)^2 + \frac{e^{2\gamma} \rho^2}{f \Delta}\dd r^2 + \frac{e^{2\gamma} \rho^2}{f}\dd \theta^2 + \frac{\Delta \sin^2 \theta}{f} \dd \phi^2\,.
\end{equation}
The functions $\rho,\Delta$ are given by
\be \rho^2 = (r-M)^2 - k^2\cos^2\theta, \qquad \Delta = r^2 - 2Mr + a^2,\ee
where $k = \sqrt{M^2-a^2}$. The event horizon is at
\be r_H = M + \sqrt{M^2-a^2}.\ee
The functions $\gamma,f,\omega$ for a generic Manko--Novikov BH depend on an infinite series of multipole parameters. We follow \cite{Staelens:2023photonrings} in only admitting an octopole-bump parameter $\alpha_3$; when $\alpha_3=0$, the metric reduces to the Kerr BH. See App.~A of \cite{Staelens:2023photonrings} for the precise expressions of $\gamma,f,\omega$ in this case.

\subsubsection{Rasheed--Larsen}
The Rasheed--Larsen BH can be obtained by a Kaluza--Klein reduction of five-dimensional Einstein gravity \cite{Rasheed:1995zv,larsen2000rotating}, i.e.\ it is a solution to the four-dimensional Lagrangian
\begin{equation}\label{eq:lagrRL}
    L = \frac{1}{16\pi G_N}\int \dd^4 x \sqrt{-g} \left[ R - 2\partial_\mu \Phi \partial^\mu \Phi - \frac14 e^{-2\sqrt{3}\Phi} F_{\mu\nu}F^{\mu\nu}\right]\,,
\end{equation}
where $R$ is the Ricci scalar, $\Phi$ is the dilaton field, and $F_{\mu\nu}$ is associated with the electromagnetic gauge field $A_\mu$.
The Rasheed--Larsen BH is a solution to Eq.~\eqref{eq:lagrRL} given by
\begin{equation}\label{eq: RL metric}
 \dd s^2 = -\frac{H_3}{\sqrt{H_1 H_2}}\left(\dd t + \boldsymbol{B}\right)^2 + \sqrt{H_1 H_2} \left(\frac{\dd r^2}{\Delta} +\dd\theta^2 + \frac{\Delta}{H_3}\sin^2\theta \dd \phi^2 \right)  \,.
\end{equation}
The functions $H_1, H_2, H_3, \Delta$, and the 1-form $\boldsymbol{B}$ are given by
\begin{align}
    H_1  = &\, r^2 + a^2 \cos^2\theta + r(p-2m) + \frac{p}{p+q}\frac{(p-2m)(q-2m)}{2} \\
  \nonumber  & - \frac{p}{2m(p+q)}\sqrt{(q^2 - 4m^2)(p^2-4m^2)} \,a \cos\theta\,, \\
    H_2 = &\,  r^2 + a^2 \cos^2\theta + r(q-2m) + \frac{q}{p+q}\frac{(p-2m)(q-2m)}{2}\\
   \nonumber &  +\frac{q}{2m(p+q)}\sqrt{(q^2 - 4m^2)(p^2-4m^2)}\, a \cos\theta\,, \\
    H_3 = &\,  r^2 + a^2 \cos^2\theta - 2mr\,, \\
    \Delta = &\,  r^2 + a^2 - 2mr\,, \\
    \boldsymbol{B} = &\, \sqrt{pq}\frac{(pq+4m^2)r - m(p-2m)(q-2m)}{2m(p+q)H_3} a\sin^2\theta \,\dd \phi\,. \label{eq: B RL}
\end{align}
The solution depends on four parameters $a, m, p, q$. These are related to the physical quantities of the metric: mass ($M$), angular momentum ($J$), and electric and magnetic charges ($Q$ and $P$) as
\begin{align}
    M & = \frac{p+q}{4}\, \label{eq: mass RL} \,,&
    J & = \frac{\sqrt{pq}(pq+4m^2)}{4m(p+q)}\,a\,, \\
    Q^2 & = \frac{q(q^2-4m^2)}{4(p+q)} \,, &
    P^2 & = \frac{p(p^2-4m^2)}{4(p+q)} \,.
\end{align}

The event horizon of the Rasheed--Larsen BH is located at
\begin{equation}
    r_H  = m + \sqrt{m^2 - a^2}\,,
\end{equation}
which implies the bound $a^2\leq m^2$. The other parameters need to satisfy $p,q \geq 2m$, where equality corresponds to the vanishing of the electric and/or magnetic charges. The case $p=q=2m$ corresponds to the chargeless Kerr limit (in which the mass is given by $m$).

\subsection{Fuzzball}\label{app:FB}
This is implemented as the {\tt ST3Cr} metric in {\tt FOORT}.\footnote{We thank Lies Van Dael for the initial implementation of this metric.} This is the ``ring fuzzball'' introduced in \cite{bacchini_fuzzball_2021}, with metric given by
\begin{equation} \label{eq:ds2multicenter} \dd s^2 = - \mathcal{Q}^{-1/2}(\dd t +  \omega)^2 + \mathcal{Q}^{1/2}\left( \dd r^2 + r^2\dd\theta^2 + r^2\sin^2\theta \dd\phi^2\right),
\end{equation}
with
\begin{equation} \mathcal{Q} = Z^3 V - \tilde\mu^2 V^2,\quad Z = L + \frac{K^2}{V},\quad \tilde\mu = \tilde M + \frac{3KL}{2V}+\frac{K^3}{V^2},\end{equation}
where the harmonic functions are given by:
\begin{align} \label{eq:harmfuncring}
\nonumber V& = \left(\frac{1}{r_1} - \frac{1}{r_2}\right), &  \tilde M &= -\frac12+\frac{P^3}{2}\left(\frac{1}{r_1} + \frac{1}{r_2}\right) + \tilde M_{D0},\\
K &= 1+P\left(\frac{1}{r_1} + \frac{1}{r_2}\right), &
L &= -P^2\left(\frac{1}{r_1} - \frac{1}{r_2}\right).
\end{align}
Here, $r_{1,2}$ are the distances on the flat three-dimensional subspace spanned by $(r,\theta,\phi)$ to the ``center'' positions $\boldsymbol{r}_{1,2}$, which are located on the $z$-axis at $z=\pm l/2$. We further have
\be\label{eq:MD0ring} \tilde M_{D0} = \frac{-2q_0}{\pi\sqrt{r^2 + R^2 +2r R \sin\theta}} \mathcal{K}\left(\frac{4r R \sin\theta}{r^2+R^2+2r R \sin\theta}\right),\ee
where $\mathcal{K}(\cdot)$ is the complete elliptical integral of the first kind,\footnote{We use the elliptical integral in terms of $m=k^2$.} 
and
\begin{align}
    l &= 8P^3 \lambda, & R &= 2\lambda\sqrt{\frac{q_0^2}{(1-(1-3P^2)\lambda)^2}-4P^6} .
\end{align}
Finally, the rotation one-form is entirely defined by
\be \label{eq:omegadiff} *_3 \dd\omega = V d\tilde M -\tilde M \dd V + \frac32\left( K \dd L - L \dd K\right),\ee
with $*_3$ the Hodge dual on the three-dimensional flat $(r,\theta,\phi)$ space; for this metric, it follows that $\omega = \omega_\phi \dd\phi$. The expression for $\omega_\phi$ in {\tt FOORT} contains an explicit integral which comes from integrating \eqref{eq:omegadiff} and makes this metric relatively computationally heavy to ray-trace.

The entire metric ultimately only depends on the parameters $P,q_0$, and $\lambda$; roughly, $P$ and $q_0$ are related to electric and magnetic charges of the solution, and $\lambda$ determines the ``size'' as well as its angular momentum. Note that these parameters cannot be chosen arbitrarily, since we must have that $l,R>0$; and moreover $P^3-q_0<0$ must hold as well for the scaling-limit ($\lambda\rightarrow0$) black hole to have a real horizon area.

This is the metric for a supersymmetric multicentered geometry or ``fuzzball''; although it derives from well-behaved branes in higher dimensions, the four-dimensional metric has naked singularities at the centers on the $z$-axis ($z=\pm l/2$) as well as the ``ring'' in the $(x,y)$ plane at $x^2+y^2=R^2$. In {\tt FOORT}, this metric implements these singularities by descending from the {\tt SingularityMetric} class which allows for singularities to be flagged in the metric. Note that the metric is axisymmetric as well as equatorially symmetric.

\subsection{Solitonic boson star}\label{sec: sol BS}

Boson stars in GR are described by the action of a complex scalar field $\varphi$ minimally coupled to gravity,
\begin{equation}
    S = \int\frac{\sqrt{-g}}{2}\left(\frac{R}{8\pi G} - \left[g^{\mu\nu}\nabla_\mu\bar{\varphi}\nabla_\nu\varphi + V(\varphi)\right]\right)\text{d}^4x\,.
\end{equation}
\FOORT provides an example of a spherically symmetric BS where the potential takes the solitonic form
\begin{equation}
    V(\varphi) = \mu_{\rm BS}^2 |\varphi|^2\left(1-2\frac{|\varphi|^2}{\sigma_0^2}\right)^2\,,
\end{equation}
with the solitonic parameter $\sigma_0 = 0.06$. The solution is obtained by assuming the harmonic ansatz $\varphi(t,r) = A(r) e^{i\omega t}$, where $r,t$ are the standard time and radial coordinates in a metric of the form
\begin{equation}
    \dd s^2 = -\alpha_{\rm BS}^2 \dd t^2 + X^2 \dd r^2 + r^2\left(\dd \theta^2 + \sin^2 \theta \dd \phi^2\right)\,.
\end{equation}
The frequency $\omega$ needs to be solved for, which is done with a shooting algorithm (not included in {\tt FOORT}) that solves the system of differential equations governing the BS model, as described in, e.g.,~Ref.~\cite{Helfer:2021brt}.
The example files in \FOORT correspond to a ground-state model (no zero crossings in $A(r)$) with central amplitude $A(0) = 0.044$ --- as considered in e.g.~Ref.~\cite{Marks:2025jpt}. This specific model is an example of an ultracompact BS, featuring a pair of light rings, one stable and one unstable. Two files are given, one containing $\Phi(r) \equiv \log (\alpha_{\rm BS}(r))$ and one containing $m(r)$ where \begin{equation}
    X^2(r) = \left(1- \frac{2m(r)}{r}\right)^{-1}\,.
\end{equation}
The BS metric is implemented using cubic-spline interpolation following Ref.~\cite{kluge-splines}. 

\section{Expressions for the fluid velocity model}\label{app: fluid velocity}

The fluid velocity model that is currently implemented in \texttt{FOORT} assumes a metric that is stationary, axisymmetric, reflection-symmetric about the equator, and circular, as well as written in coordinates adapted to circularity:
\begin{itemize}
    \item Stationarity and axisymmetry are assumed as $p_t,p_\phi$ are implicitly assumed to be conserved quantities for the fluid trajectory geodesics.
    \item Equatorial-reflection symmetry is needed to ensure that geodesic trajectories confined to the equatorial plane exist.\footnote{Note that the Manko-Novikov black hole imaged in Fig. \ref{fig: MN FB} does not satisfy equatorial reflection symmetry. However, it can be shown that the ``true disc location'' (the $\theta$ angle at which a constant-$r$ orbit can exist) for this black hole is not far from the equatorial plane $\theta=\pi/2$, see Appendix D.2 of \cite{Staelens:2023photonrings}.}
    \item The fluid velocity model also assumes that the metric is diagonal in $r$ and $\theta$ and block-diagonal in $(t,\phi)$ (in particular, metric components such as $g_{tr},g_{r\phi}$ must vanish). Circular metrics can always be written in such a form.
\end{itemize}
In addition, asymptotic flatness is assumed when calculating the radially infalling velocity (see below in App.~\ref{app: radial fluid velocity}).

Care must be taken when using this fluid velocity model on a metric that does not satisfy these constraints, as undefined and unphysical behavior may result. For example, using this fluid model on the Kerr--Schild form~\eqref{eq: Kerr metric 2} of the Kerr metric (which e.g.\ has $g_{tr}\neq 0$) gives different (and wrong) emission compared to the Kerr metric. In certain cases (especially when e.g.\ equatorial symmetry is broken slightly), the fluid model can still give sensible results --- see e.g.\ the Manko--Novikov picture of Fig.~\ref{fig: MN FB}.

\subsection{Finding the ISCO}

We briefly detail how the ISCO is determined in {\tt FOORT}. We take it to be the outermost radius at which stable circular orbits (at larger radii) give way to unstable or non-existent ones (at smaller radii). For BHs such as Kerr this is the usual ISCO; for horizonless objects with an inner region of stable circular orbits, such as the BS of Sec. \ref{sec: demos}, it is the outer marginally stable circular orbit (MSCO). The search is confined to the equatorial plane $\theta=\pi/2$ and to an interval $[R_L, R_U]$\footnote{For a metric with a horizon at $r=r_h$, {\tt FOORT} initially specifies $R_L =r_h$ and $R_U=10r_h$; for other metrics, {\tt FOORT} uses the defaults $R_L=0.05$ and $R_U=1000$, unless otherwise specified.} and proceeds as follows:
\begin{itemize}
    \item Check that a stable circular orbit exists at $R_U$; otherwise, no ISCO is determined.
    \item Scan inward from $R_U$ in 200 equal steps until a radius is found without a stable circular orbit (i.e. the circular orbit is unstable, or no circular orbit exists). If there is no such radius, no ISCO is determined.
    \item Bisect between this (unstable circular orbit) radius and the last stable orbit to further narrow down the ISCO location. We repeat the bisection until a specified tolerance is reached, or a maximal number of iterations has taken place.
\end{itemize}
A warning is printed if the result lies within $1\%$ of the search interval of either bound.

\subsection{Circular velocity outside the ISCO}

Consider an equatorial, circular timelike geodesic with conserved momenta $p_t=-E$ and $p_\phi=L$, and $p_r=p_\theta=0$, parametrized by $\lambda$. The radial geodesic equation reads
\begin{equation}
  \frac{\dd^2 r}{\dd \lambda^2} = -F(r), \qquad F(r) \equiv \Gamma^{r}{}_{\mu\nu} \, g^{\mu a}\, p_a\, g^{\nu b}\, p_b,
\end{equation}
and a circular orbit at radius $r_0$ requires $F(r_0)=0$. Defining
$\mathcal{A}=\Gamma^r{}_{\mu\nu}g^{\mu t}g^{\nu t}$,
$\mathcal{B}=\Gamma^r{}_{\mu\nu}g^{\mu\phi}g^{\nu\phi}$, and
$\mathcal{C}=2\,\Gamma^r{}_{\mu\nu}g^{\mu t}g^{\nu\phi}$, this condition becomes a quadratic equation for $\eta \equiv L/E$,
\begin{equation}
  \mathcal{B}\,\eta^2 - \mathcal{C}\,\eta + \mathcal{A} = 0
  \quad\Rightarrow\quad
  \eta = \frac{\mathcal{C} \pm \sqrt{\mathcal{C}^2 - 4\mathcal{A}\mathcal{B}}}{2\mathcal{B}} ,
\end{equation}
where we select the root\footnote{We evaluate the two roots in the numerically stable form $\eta_1=q/\mathcal{B}$, $\eta_2=\mathcal{A}/q$ with $q=\tfrac12\big[\mathcal{C}+\operatorname{sgn}(\mathcal{C})\sqrt{\mathcal{C}^2-4\mathcal{A}\mathcal{B}}\big]$. As $\mathcal{B}\to0$, only the finite root $\mathcal{A}/\mathcal{C}$ remains; if $\mathcal{C}$ vanishes as well, as happens in static spacetimes wherever $\mathcal{B}=0$ (e.g.\ at a horizon or at an extremum of $g_{\phi\phi}$), no circular orbit exists at that radius and the fluid follows the radially infalling prescription of Appendix~\ref{app: radial fluid velocity}.} that maximizes $\eta$ while keeping the geodesic timelike, corresponding to prograde orbits when the angular momentum of the spacetime is positive.\footnote{As mentioned, this fluid velocity model assumes axisymmetry, which implies that the angular momentum ($J_\phi$) of the spacetime is well-defined and constant. Note that for $J_\phi<0$, the orbit of the fluid is retrograde.}
The four-velocity normalization $g^{ab}p_ap_b=-1$ then fixes the energy,
\begin{equation}
  E^2 = \left(-g^{tt} + 2g^{t\phi}\eta - g^{\phi\phi}\eta^2\right)^{-1}\,.
\end{equation}
Thus, a timelike circular orbit exists at $r_0$ if and only if $\mathcal{C}^2 \geq 4\mathcal{A}\mathcal{B}$ and the bracket above is positive. We end up with
\begin{align}
    -p_0 & = E =\left(-g^{tt} + 2g^{t\phi}\eta - g^{\phi\phi} \eta^2 \right)^{-1/2}\,, \\
    p_\phi & = \eta E,
\end{align}
and
\begin{equation}
    \left(p_{\text{circ}}\right)_\mu  = \left(p_0, 0, 0, p_\phi\right)\,.
\end{equation}
To make this sub-Keplerian, we reduce the angular momentum with $\xi \in [0,1]$ as $p_\phi \to \xi p_\phi$. This then warrants a recalculation of $p_0$ \cite{Penna:2013zga}.

\subsubsection{Stability of circular geodesics}\label{app: circular orbit stability}

To assess the stability of the circular orbit, we perturb the radius, $r = r_0 + \delta r$, keeping $E$ and $L$ fixed. To linear order, this gives
\begin{equation}
  \frac{\dd^2 \delta r}{\dd \lambda^2} = -F'(r_0)\,\delta r ,\qquad
  F'(r_0) = \partial_r\!\left(\Gamma^r{}_{\mu\nu}g^{\mu a}g^{\nu b}\right)\!\bigg|_{r_0}\,p_a\,p_b ,
\end{equation}
such that the orbit is stable if $F'(r_0)>0$. The radial derivative is evaluated using finite differencing with a fourth-order central stencil.

\subsection{Circular velocity inside the ISCO}

We extend the circular flow smoothly inside the ISCO according to Cunningham's prescription \cite{Cunningham1975}. The conserved energy and angular momentum are assumed to be those of the geodesic associated with the ISCO; the four-velocity then requires a non-zero component for the radial momentum such that the normalization $p_\mu p^\mu =  -1$ is satisfied. In the case that unstable circular orbits exist inside the ISCO, in particular, this means that the momentum given by the above prescription does \textit{not} agree with that of the unstable circular geodesic.

If extending the flow in this way does not succeed, and a circular velocity cannot be found, {\tt FOORT} sets the fluid velocity to be purely radially infalling. This can happen, e.g., for horizonless spacetimes, such as the BS, if Cunningham's prescription is applied from an MSCO inwards (see Sec.~\ref{sec: demos}). Towards the center of the star, a centrifugal barrier is hit beyond which the conserved quantities of the MSCO do not allow for a circular-infalling velocity.

\subsection{Radially infalling velocity}\label{app: radial fluid velocity}

The fluid velocity for radially infalling particles is easily determined by relating it to a geodesic that is radially infalling from infinity, with zero initial velocity.\footnote{Note that this implicitly assumes the metric is asymptotically flat.}
For the latter, at infinity, we have
\begin{equation}
    \left(u_{\text{rad}}\right)_\mu = (-E, 0,0,0)\,, \quad E = 1\,.
\end{equation}
At a finite radius $r$, the normalization $p_\mu p^\mu = -1$ implies that
\begin{equation}
    g^{tt} - 2g^{tr}p_r + g^{rr} p_r^2 = -1\,,
\end{equation}
where we leave the possibility for a $tr$-component in the metric. From the above, the expression for $p_r$ follows, where the negative sign makes the fluid infalling:
\begin{equation}
    p_r = \frac{1}{g^{rr}}\left( g^{tr} - \sqrt{(g^{tr})^2 - g^{rr} \left(g^{tt} + 1\right)}\right)\,.
\end{equation}

\printbibliography

\end{document}